\documentclass[final,3pt]{elsarticle}

\usepackage{lineno,hyperref}
\usepackage{amssymb,amsmath}
\usepackage{mathtools}
\usepackage{longtable}
\usepackage{booktabs}
\newcommand*\rfrac[2]{{}^{#1}\!/_{#2}}
\usepackage[overload]{empheq}
\usepackage{afterpage}
\usepackage{xcolor}
\usepackage{tikz}
\usetikzlibrary{calc,trees,positioning,arrows,chains,shapes.geometric,%
decorations.pathreplacing,decorations.pathmorphing,shapes,%
matrix,shapes.symbols,plotmarks,decorations.markings,shadows}
\usetikzlibrary{chains, fit, positioning}
\usepackage{stmaryrd}
\usepackage{bm}
\usepackage[shortlabels]{enumitem}
\usepackage{pgfplots}
\usepackage{subcaption}
  \pgfplotsset{compat=newest}
  \usetikzlibrary{plotmarks}
  \usetikzlibrary{arrows.meta}
  \usepgfplotslibrary{patchplots}
  \usepackage{grffile}

\usepackage{geometry}

\modulolinenumbers[5]

\newenvironment{review}{\color{black}}{}
\newcommand{\rev}[1]{\begin{review}#1\end{review}}

\newenvironment{revieww}{\color{black}}{}
\newcommand{\revv}[1]{\begin{revieww}#1\end{revieww}}

\DeclarePairedDelimiterX{\infdivx}[2]{(}{)}{%
  #1\;\delimsize\|\;#2%
}

\DeclarePairedDelimiter{\norm}{\lVert}{\rVert}

\journal{Mechanical Systems and Signal Processing}

\begin{document}

\begin{frontmatter}

\title{Retrieving highly structured models starting from black-box nonlinear state-space models using polynomial decoupling}


\author[label1,label2]{J. Decuyper}
\ead{jan.decuyper@vub.be}
\author[label3]{K. Tiels}
\author[label1,label2]{M. C. Runacres}
\author[label1,label4]{J. Schoukens}


\address[label1]{Vrije Universiteit Brussel (VUB), Department of Engineering Technology (INDI), Pleinlaan 2, 1050 Brussels, Belgium.}
\address[label2]{Vrije Universiteit Brussel (VUB), Thermo and Fluid Dynamics (FLOW), Pleinlaan 2, 1050 Brussels, Belgium.}
\address[label3]{Uppsala University (UU), Department of Information Technology, PO Box 337, SE-75105 Uppsala, Sweden.}
\address[label4]{Eindhoven University of Technology (TU/e), Department of Electrical Engineering, Eindhoven, The Netherlands.}

%
%

\begin{abstract}

Nonlinear state-space modelling is a very powerful black-box modelling approach. However powerful, the resulting models tend to be complex, described by a large number of parameters. In many cases interpretability is preferred over complexity, making too complex models unfit or undesired. In this work, the complexity of such models is reduced by retrieving a more structured, parsimonious model from the data, without exploiting physical knowledge. Essential to the method is a translation of all multivariate nonlinear functions, typically found in nonlinear state-space models, into sets of univariate nonlinear functions. The latter is computed from a tensor decomposition. It is shown that typically an excess of degrees of freedom are used in the description of the nonlinear system whereas reduced representations can be found. The method yields highly structured state-space models where the nonlinearity is contained in as little as a single univariate function, with limited loss of performance. Results are illustrated on simulations and experiments for: the forced Duffing oscillator, the forced Van der Pol oscillator, a Bouc-Wen hysteretic system, and a Li-Ion battery model.

\end{abstract}


\end{frontmatter}


\section{Introduction}
\label{s:Intro}



For a large number of engineering applications black-box models are essential tools when describing systems. Often white-box models are prohibitively expensive or too complex to be derived from physics. Under these conditions data-driven modelling can provide an efficient alternative. The model structures, suitable for black-box identification, are designed to be generic, i.e.\ they allow for a large number of degrees of freedom. Two downsides of such models are that they typically require a large number of parameters and that the nonlinear elements appear `unstructured' in the equations. In this setting unstructured refers to the fact that large multivariate nonlinear functions are introduced. Having multivariate (also cross-term) nonlinear parts hinders the interpretability of the model and by extension of the system.

A solution to limiting the dimensions of the identified model is by applying model selection/reduction. This can be done prior to the identification procedure, by encoding known properties of the system directly into the structure, e.g.\ using a distortion analysis to discriminate between odd or even nonlinear behaviour \cite{pintelon2001}. But also during the identification itself the necessity of an additional degree of freedom can be evaluated. In \cite{bai2014} a variable selection algorithm is proposed for the identification of NARX models \cite{billings2013}, thereby tackling the curse of dimensionality. In this work we will focus on an a posteriori model reduction, with the aim of removing redundant degrees of freedom and retrieving highly structured nonlinear representations, i.e.\ univariate nonlinear functions, without a loss of accuracy of the model.

The decoupling of multivariate polynomials was already studied in the context of system identification. In \cite{usevich2014,schoukensM2012} it was shown that the cross-terms of a multivariate polynomial can be eliminated by diagonalising the matrix holding the coefficients of the polynomial. The method, however, resulted in a tensor decomposition of which the order of the tensor grows with the degree of the polynomial.

An alternative decoupling technique was proposed in \cite{dreesen2014}. It was shown that decoupling the multivariate polynomial into univariate functions boils down to introducing appropriate linear transformations of the inputs and the outputs of the coupled function. The latter is inferred from the first order derivative information of the coupled function and involves decomposing a three-way tensor, irrespective of the degree of the polynomial. A revision of the method is provided in Section \ref{ss:decoupling}. 

Also NARX models suffer from a combinatorial growth of the number of parameters, both with the maximum lags and the maximum polynomial degree. \begin{review} In that case the multivariate function has a single output, resulting in a Jacobian matrix (first order derivative) rather than a tensor. Not being able to exploit the uniqueness properties of tensor decomposition introduces significant complications \cite{decuyper2019}. The issue arising from non unique tensor decomposition is described in detail in Section \ref{ss:remarks} and an overview of the existing solutions is presented in Section \ref{ss:avoiding_pr}.\end{review}
 
In this paper we focus on multivariate polynomials that appear in polynomial nonlinear state-space models (PNLSS) \cite{paduart2010}. Enabling static as well as dynamic nonlinearities, both in feedforward and feedback, this model class covers a wide variety of nonlinear systems. Models were successfully identified for an hysteresis system in \cite{noel2017}, a Li-ion battery during a loading cycle in \cite{relan2016} and unsteady fluid dynamics in \cite{decuyper2018}. The issue of over-parameterisation of such models was discussed in \cite{young2018}.

Recent results of applying polynomial decoupling to PNLSS models were presented for the Bouc-Wen hysteresis system in \cite{fakhrizadeh2018,dreesen2016}. The authors were able to reduce the number of nonlinear parameters from 90 for the full PNLSS model to 51 for the decoupled representation. The reduced model, moreover, yielded a decreased validation error, highlighting the hazard of local minima when identifying the large PNLSS models. 

In this work the decoupling technique is extended with a \begin{review} number of additional \end{review} model reduction steps. An overview of the contributions is provided next.

\subsection{Contributions presented in this work}
\label{ss:contr}

In this work we present reduced nonlinear state-space models for the following systems: the forced Duffing oscillator, a Bouc-Wen hysteretic system, the forced Van der Pol equation, and a Li-Ion battery. The proposed reduction follows from \rev{progressively} reshaping the nonlinear functions. During this procedure successive reduction steps are applied while the accuracy of the model is monitored. As the reduction proceeds, the functions evolve towards their final form. Three actions are proposed:
\begin{enumerate}
\item Decouple the multivariate nonlinear function into univariate functions of intermediate variables (Section \ref{ss:decoupling}).
\item Approximate the decoupled function using only identical univariate functions over multiple branches (Section \ref{ss:unifying}).
\item Exploit the fact that the branches are generally not linearly independent to reduce their number (Section \ref{ss:branch_reduction}).
\end{enumerate}

The method decreases the complexity of the model through an efficient parameterisation of the nonlinear part while maintaining high accuracy via repeated optimisation steps. This results in highly structured models with very compact formulations of the nonlinearity.

\subsection{A motivating example}
\label{ss:motivating_ex}
\begin{figure}
\begin{subfigure}[b]{0.45\textwidth}
\begin{center}
\includegraphics[width=\textwidth]{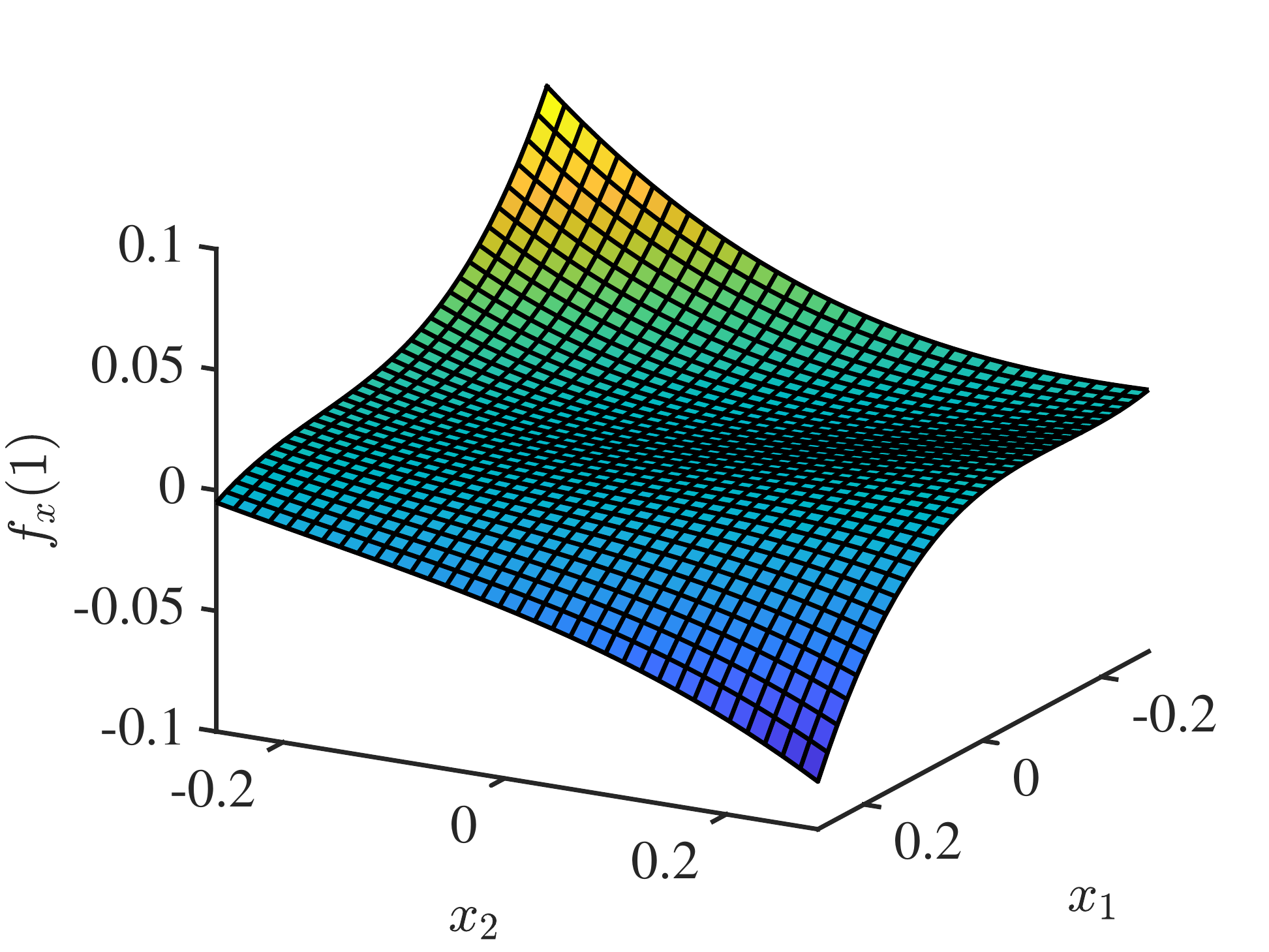}
\caption{}
\end{center}
\end{subfigure}
\begin{subfigure}[b]{0.45\textwidth}
\begin{center}
\includegraphics[width=\textwidth]{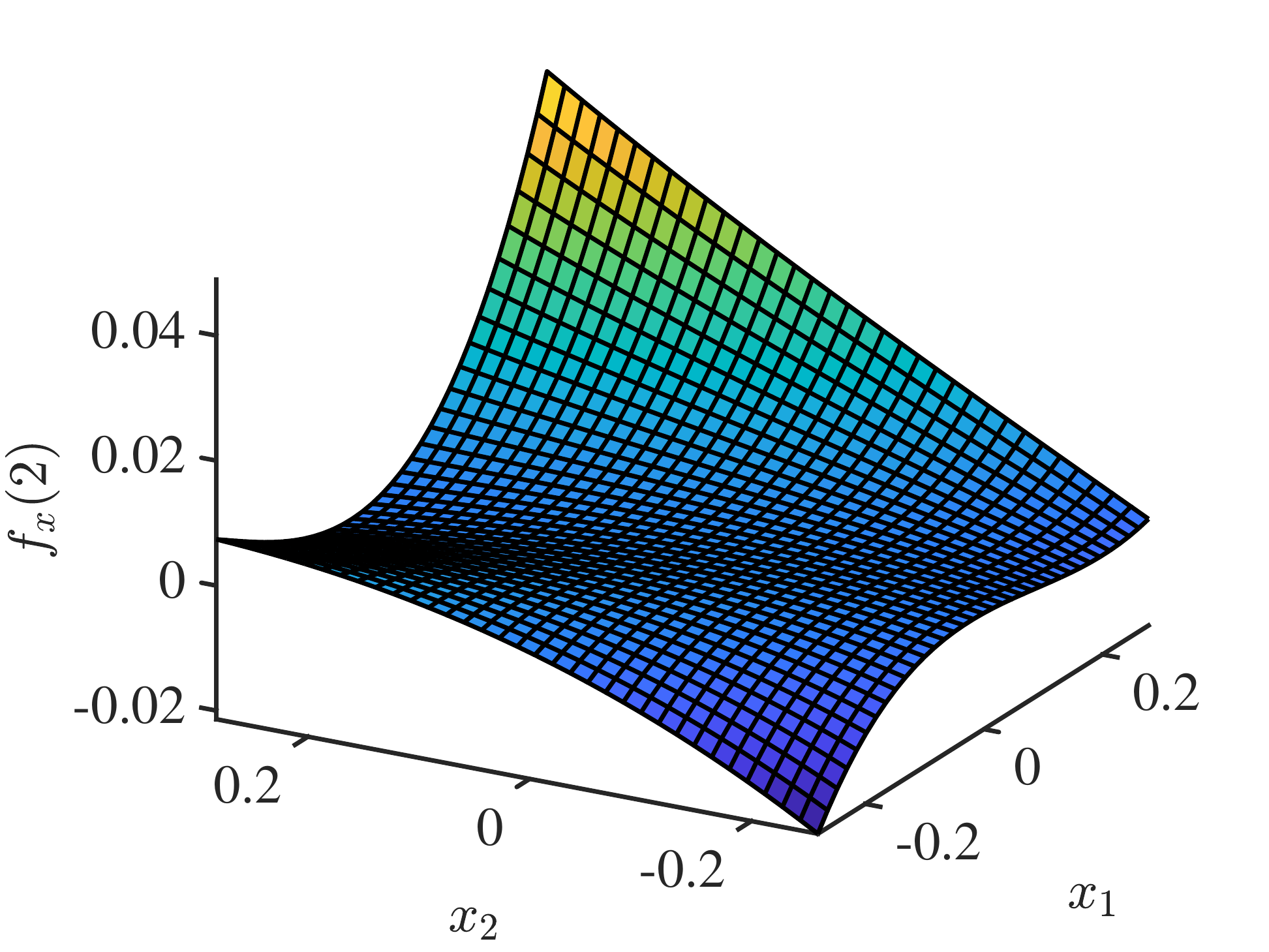}
\caption{}
\end{center}
\end{subfigure}
\caption{Visualisation of the coupled multivariate polynomial of the forced Duffing model (Eq.~\eqref{e:SB1}) over the operating regime. Panel (a) and (b) show both elements of the vector function $\textbf{f}_x$ (Eq.~\eqref{f:fx_fyA} where $\textbf{f}_x$ does not depend on $\textbf{u}$ in the case of the forced Duffing model).}
\label{f:SB1}
\end{figure}

%
\revv{The present work achieves a complexity reduction of nonlinear state-space models by means of reducing the static nonlinear functions present in the model. Here, we illustrate this on a model of the forced Duffing oscillator \cite{wigren2013} (this is discussed in detail in Section \ref{ss:silverbox}). Omitting the dynamic part of the model, for now, (formal definitions are introduced in Section \ref{ss:def_PNLSS}) we are able to illustrate the reduction that can be achieved. Eq~\eqref{e:SB11} represents the nonlinear part in the state update of this model.%
\begin{equation}
\textbf{f}_x(\textbf{x}) =  \left[ \begin{matrix} e_{11} \quad e_{12} \quad e_{13} \quad e_{14} \quad e_{15} \quad e_{16} \quad e_{17}  \\ e_{21} \quad e_{22} \quad e_{23} \quad e_{24} \quad e_{25} \quad e_{26}\quad e_{27} \end{matrix} \right]
       \left[ \begin{matrix} x_1^2(k) \\ x_1(k)x_2(k) \\ x_2^2(k) \\ x_1^3(k) \\ x_1^2(k)x_2(k) \\ x_1(k)x_2^2(k) \\ x_2^3(k) \end{matrix} \right].
       \label{e:SB11}
 \end{equation}
 It's a static vector function with two inputs, $x_1$ and $x_2$, \revv{and $e_{ij}$ the polynomial coefficients that were estimated from data. The function contains coupled monomial terms (cross products) of the second and the third degree.} Both function components are represented graphically in Fig.~\ref{f:SB1}. The coupled terms make the nonlinearity hard to grasp. Moreover, 14 parameters, $e_{ij}$, are used in the description of the function.}
 
\begin{figure}
\begin{center}
\includegraphics[width=0.45\textwidth]{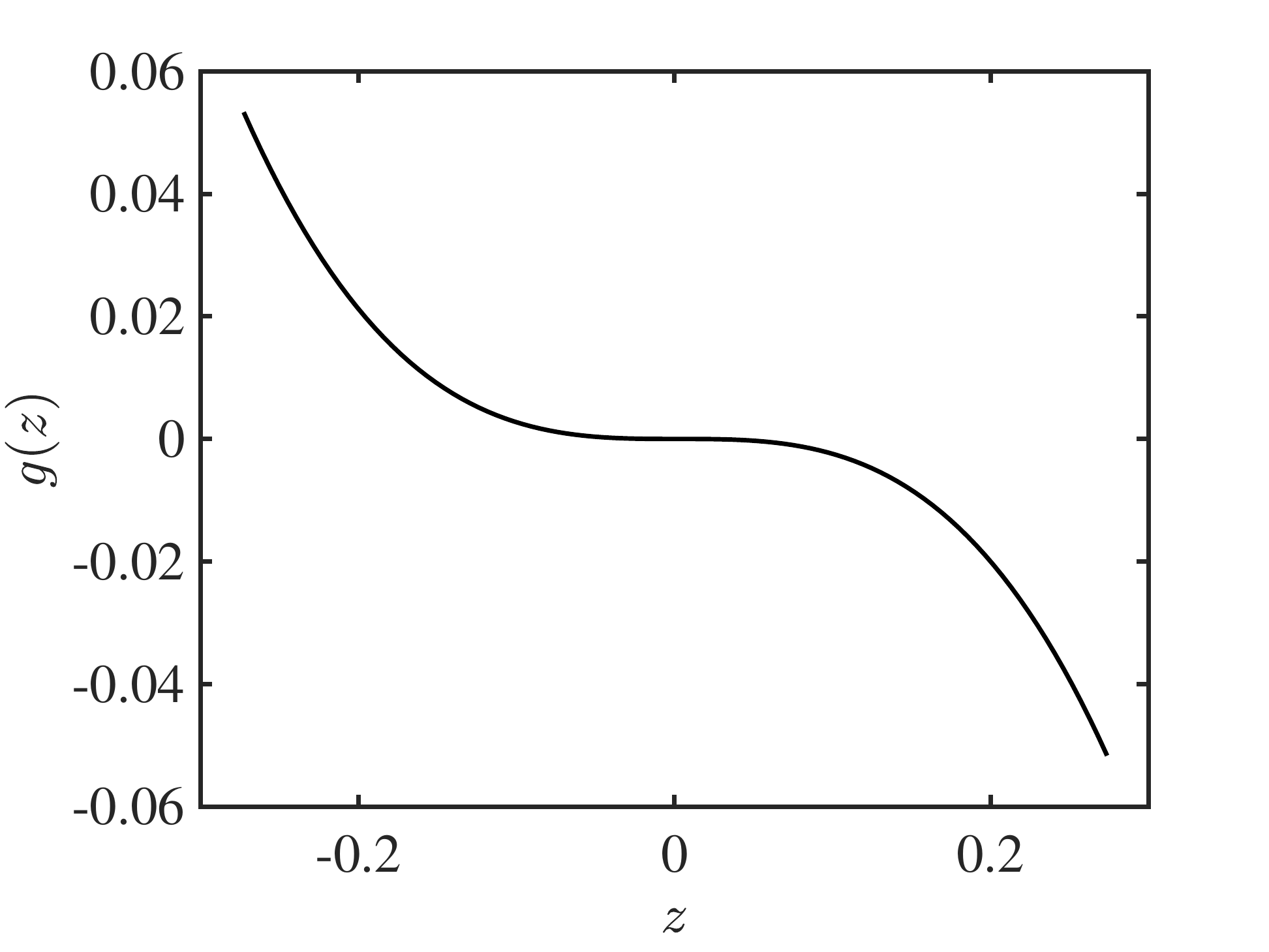}
\caption{Visualisation of the decoupled univariate polynomial of the forced Duffing model when reduced to a single branch. The univariate function $g$ is the nonlinear element of the decoupled representation of Eq.~\ref{e:dec1}.}
\label{f:SB2}
\end{center}
\end{figure}

 \revv{Without any loss of accuracy, a reduced model is derived where in this case the static nonlinear part is of the following form:
\begin{subequations} \label{e:SB2}
\begin{align}
 &\textbf{f}_x =  \left[ \begin{matrix} w_1  \\ w_2  \end{matrix} \right] \overbrace{\left[ \theta_{1}z^3(k)+ \theta_{2}z^2(k)\right]}^{g(z)} \\
 &z(k) = [v_1 \quad v_2] \left[ \begin{matrix} x_1(k) \\ x_2(k) \end{matrix} \right].
 \end{align}
 \end{subequations}
 The nonlinearity is in this case contained in a single univariate polynomial function $g$, using a linear combination of the inputs as intermediate variable. A univariate function is much more tractable than a multivariate one. A visualisation of $g$ is presented in Fig.~\ref{f:SB2}. Notice that this description requires only 6 parameters.}

\subsection{Layout}

Section \ref{s:retrieving_struct} revises the method of polynomial decoupling from first order information. Reshaping the \revv{decoupled} function, either via imposing identical branches or by reducing the number of branches is the subject of Section \ref{s:reduction}. In Section \ref{s:case_studies} the proposed model reduction method is illustrated on a number of numerical and experimental case studies. Section \ref{s:conclusions} provides some conclusions.

\subsection{Notation}
Vectors are denoted by lowercase bold-faced letters, e.g.\ $\textbf{x} \in \mathbb{R}^{n}$. Matrices are given bold-faced uppercase letters, e.g.\ \textbf{A} with its columns $(\textbf{a}_1, \dots, \textbf{a}_n) \in \mathbb{R}^{n \times n}$. Tensors are denoted by caligraphic letters, e.g.\ $\mathcal{T} \in \mathbb{R}^{n \times m \times N}$. Time derivatives are denoted by overdots, \revv{$\frac{d\textbf{a}(t)}{dt}= \dot{\textbf{a}}(t)$}. 

\section{Prerequisites on polynomial decoupling using tensor methods}
\label{s:retrieving_struct}

We will address multivariate polynomials in the context of nonlinear state-space models. 

\subsection{Discrete-time polynomial nonlinear state-space models}
\label{ss:def_PNLSS}
The generic set of equations of such models is given by
\begin{subequations} \label{e:PNLSS1}
    \begin{empheq}[left={\empheqlbrace\,}]{align}
      & \textbf{x}(k+1)=\textbf{A}\textbf{x}(k)+\textbf{B}\textbf{u}(k)+\textbf{E}\bm{\zeta}(\textbf{x}(k),\textbf{u}(k)) \label{e:PNLSS1_a} \\ 
      & \textbf{y}(k)=\textbf{C}\textbf{x}(k)+\textbf{D}\textbf{u}(k)+\textbf{F}\bm{\eta}(\textbf{x}(k),\textbf{u}(k)) \label{e:PNLSS1_b},
    \end{empheq}
\end{subequations}
where $k=\rfrac{t}{T_s}$ is the time index with $T_s$ the sampling period. \revv{The model is described by $n$ state variables (model order), $\textbf{x}(k) \in \mathbb{R}^{n}$, $p$ outputs, $\textbf{y}(k) \in \mathbb{R}^p$, and $m$ inputs, $\textbf{u}(k) \in \mathbb{R}^m$. The matrices then have the following dimensions: $\textbf{A} \in \mathbb{R}^{n \times n}$,  $\textbf{B} \in \mathbb{R}^{n \times m}$,  $\textbf{C} \in \mathbb{R}^{p \times n}$,  $\textbf{D} \in \mathbb{R}^{p \times m}$, $\textbf{E} \in \mathbb{R}^{n \times n_{\zeta}}$, and $\textbf{F} \in \mathbb{R}^{p \times n_{\eta}}$.} When $\bm{\zeta}$ and $\bm{\eta}$ contain multivariate monomials this class of models is known as the discrete-time polynomial nonlinear state-space models (PNLSS) \cite{paduart2010}. \revv{The vectors of monomials may contain up to all possible cross products between states and input variables raised to some user defined powers.} For notational convenience the multivariate polynomial function in the state equation will be referred to with $\textbf{f}_x(\textbf{x},\textbf{u})$ while the nonlinear function in the output equation will be denoted $\textbf{f}_y(\textbf{x},\textbf{u})$,
\begin{subequations}
\label{f:fx_fy}
\begin{align}
\textbf{f}_x(\textbf{x},\textbf{u}) &= \textbf{E}\bm{\zeta}(\textbf{x}(k),\textbf{u}(k)), \label{f:fx_fyA}\\
\textbf{f}_y(\textbf{x},\textbf{u}) &= \textbf{F}\bm{\eta}(\textbf{x}(k),\textbf{u}(k)).\label{f:fx_fyB}
\end{align}
\end{subequations}

Note that the number of parameters in these state-space models is larger than the number of degrees of freedom (DOF).
For example, a linear state transformation $\textbf{x}(k) = \textbf{T} \textbf{x}_T(k)$ could be introduced, where \textbf{T} is an arbitrary invertible matrix. Introducing this state transformation in Eq.~\eqref{e:PNLSS1} leads to a state-space representation
\begin{subequations}
\label{e:PNLSS_T}
\begin{align}
	\textbf{x}_T(k+1) 	&= \textbf{T}^{-1} \textbf{A} \textbf{T} \textbf{x}_T(k) + \textbf{T}^{-1} \textbf{B} \textbf{u}(k) + \textbf{T}^{-1} \textbf{E} \bm{\zeta}(\textbf{T} \textbf{x}_T(k),\textbf{u}(k))\\
	\textbf{y}(k)				&= \textbf{C} \textbf{T} \textbf{x}_T(k) + \textbf{D} \textbf{u}(k) + \textbf{F} \bm{\eta}(\textbf{T} \textbf{x}_T(k),\textbf{u}(k))
\end{align}
\end{subequations}

The state transformation does not change the input/output behaviour of the state-space model, it only changes the parameterisation.
If $n$ is the size of the state vector, $n^2$ elements in \textbf{T} can be chosen freely (as long as \textbf{T} is invertible). Hence the actual number of DOF is in the $\mathcal{O}(n^2)$ smaller than the number of parameters in the fully parameterised state-space model (where the number of parameters is the sum of the number of elements in all the state-space matrices).
In the remainder of this paper, we will report the number of DOF instead of the number of parameters.
The number of DOF is computed as the rank of the Jacobian matrix, which collects the partial derivatives of the outputs \textbf{y} with respect to the parameters.

Models of the form of Eq.~\eqref{e:PNLSS1} are used for simulation purposes. Their accuracy is measured from the least squares cost on the simulated output,
\begin{equation}
\label{e:VLS}
\textbf{V}_{\text{LS}}(\bm{\theta}_{ss}) =  \frac{1}{N}\sum_{k=1}^N  \norm{\textbf{y}_{\text{meas}}(k) - \textbf{y}(\bm{\theta}_{ss},k)}^2_2,
\end{equation}
where $\textbf{y}_{\text{meas}}$ is the vector of measured outputs, $\textbf{y}$ is the vector of simulated outputs, $N$ is the number of samples, and $\bm{\theta}_{ss}$ is a vector collecting all the model parameters.


\subsection{Decoupling multivariate polynomials from first order information}
\label{ss:decoupling}
The present section is based predominantly on the work of Dreesen et al.\ \cite{dreesen2014}.

\begin{figure}[h]
\begin{center}

\setlength{\unitlength}{1cm}
\begin{tikzpicture}[scale=0.6] 


\node at (-0.5,4.5) {\small{$x_1$}};
\node at (-0.5,3) {\small{$x_n$}};

\node at (-0.5,2) {\small{$u_1$}};
\node at (-0.5,0.5) {\small{$u_m$}};
\draw [->,thick] (0,4.5) -- (1,4.5);
\draw[dotted,thick] (0.5,3.55) -- (0.5,3.95);
\draw [->,thick] (0,3) -- (1,3);
\draw [->,thick] (0,2) -- (1,2);
\draw[dotted,thick] (0.5,1.05) -- (0.5,1.45);
\draw [->,thick] (0,0.5) -- (1,0.5);

\draw[thick,rounded corners, fill=lightgray] (1,2) -- (1,5) -- (7,5) -- (7,0) -- (1,0) -- (1,2);


\node at (4,2.5) {\small{$\textbf{f}_x(\textbf{x},\textbf{u})$}};

\draw [->,thick] (7,4) -- (8,4);
\draw[dotted,thick] (7.5,2.4) -- (7.5,2.8);
\draw [->,thick] (7,1) -- (8,1);

\node at (8.6,4) {\small{$q_1$}};
\node at (8.6,1) {\small{$q_n$}};

\draw[<->, ultra thick] (10,2.5) -- (11.5,2.5);


\node at (13,4.5) {\small{$x_1$}};
\node at (13,3) {\small{$x_n$}};

\node at (13,2) {\small{$u_1$}};
\node at (13,0.5) {\small{$u_m$}};

\draw [->,thick] (13.5,4.5) -- (14.5,4.5);
\draw[dotted,thick] (14,3.55) -- (14,3.95);
\draw [->,thick] (13.5,3) -- (14.5,3);
\draw [->,thick] (13.5,2) -- (14.5,2);
\draw[dotted,thick] (14,1.05) -- (14,1.45);
\draw [->,thick] (13.5,0.5) -- (14.5,0.5);

\draw [thick] (14.5,0) rectangle(16,5);
\node at (15.25,2.5) {$\textbf{V}_x^{\text{T}}$};

\draw [->,thick] (16,4) -- (17.5,4);
\draw[dotted,thick] (18.5,2.4) -- (18.5,2.8);
\draw [->,thick] (16,1) -- (17.5,1);



\draw[thick,rounded corners, fill=lightgray] (17.5,4) -- (17.5,4.475) -- (19.5,4.475) -- (19.5,3.525) -- (17.5,3.525) -- (17.5,4);
\node at (18.5,4) {\footnotesize{$g_{x_1}$}\footnotesize{$(z_1)$}};

\draw[thick,rounded corners, fill=lightgray] (17.5,1) -- (17.5,1.475) -- (19.5,1.475) -- (19.5,0.525) -- (17.5,0.525) -- (17.5,1);
\node at (18.5,1) {\footnotesize{$g_{x_r}$}\footnotesize{$(z_r)$}};

\draw [->,thick] (19.5,4) -- (21,4);
\draw [->,thick] (19.5,1) -- (21,1);

\draw [thick] (21,0) rectangle(22.5,5);
\node at (21.75,2.5) {$\textbf{W}_x$};

\draw [->,thick] (22.5,4) -- (23.5,4);
\draw[dotted,thick] (23,2.4) -- (23,2.8);
\draw [->,thick] (22.5,1) -- (23.5,1);

\node at (24.1,4) {\small{$q_1$}};
\node at (24.1,1) {\small{$q_n$}};

\end{tikzpicture}
\caption{Graphical illustration of the coupled multivariate nonlinear function $\textbf{f}_x$ on the left hand side and a decoupled formulation with nonlinear univariate branches $\textbf{g}_x$ on the right hand side. A similar representation can be written for $\textbf{f}_y$.}
\label{f:decoupling}
\end{center}
\end{figure}
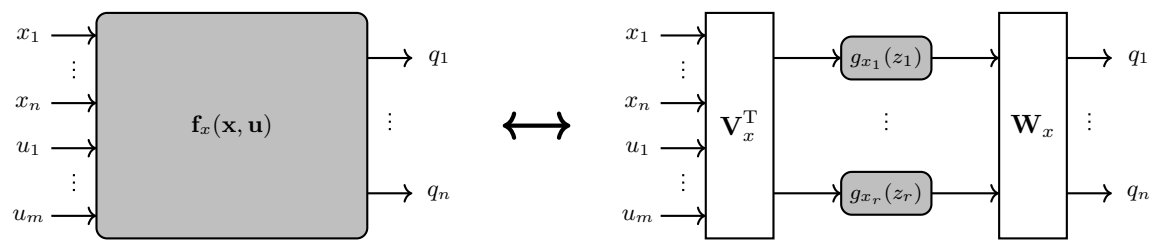

Consider the coupled multivariate polynomial $\textbf{f}_x(\textbf{x},\textbf{u})$ of \revv{Eq.~\eqref{f:fx_fyA}}, with $n+m$ inputs, resulting respectively from the dimension of the state vector and the input vector of the state-space model. The number of outputs of $\textbf{f}_x$ then equals $n$. Following \cite{dreesen2014} we say that $\textbf{f}_x$ has a decoupled representation when it can be written as
\begin{equation}
\label{e:dec1}
\textbf{f}_x(\textbf{x},\textbf{u}) = \textbf{W}_x \textbf{g}_x\left(\textbf{V}_x^{\text{T}} \left[ \begin{matrix} \textbf{x} \\ \textbf{u} \end{matrix} \right]\right),
\end{equation}
where for notational convenience the reference to time is omitted. Eq.~\eqref{e:dec1} is decoupled in the sense that the functions $\textbf{g}_x: \mathbb{R}^{r_x} \rightarrow \mathbb{R}^{r_x}$ \revv{has $r_x$ components where the $i$-th component function is a univariate polynomial of the $i$-th intermediate variable}. The intermediate variables follow from a linear transformation $\textbf{V}_x = (\textbf{v}_{x_1}, \dots, \textbf{v}_{x_{r_x}}) \in \mathbb{R}^{(n+m) \times r_x}$ of the inputs \textbf{x} and \textbf{u}. A second linear transformation $\textbf{W}_x = (\textbf{w}_{x_1}, \dots, \textbf{w}_{x_{r_x}}) \in \mathbb{R}^{n \times r_x}$, applied to the outputs of $\textbf{g}_x$ completes the formulation. A graphical illustration of a coupled versus a decoupled multiple-input multiple-output function is provided in Fig.~\ref{f:decoupling}. 

Analogously also the multivariate output nonlinearity, $\textbf{f}_y(\textbf{x},\textbf{u})$, is parametrised using univariate functions $\textbf{g}_y$. The resulting state-space model is of the form
\begin{subequations} \label{e:PNLSS2}
    \begin{empheq}[left={\empheqlbrace\,}]{align}
      & \textbf{x}(k+1)=\textbf{A}\textbf{x}(k)+\textbf{B}\textbf{u}(k)+ \textbf{W}_x \textbf{g}_x\left(\textbf{V}_x^{\text{T}} \left[ \begin{matrix} \textbf{x} \\ \textbf{u} \end{matrix} \right]\right) \label{e:PNLSS2_a} \\ 
      & \textbf{y}(k)=\textbf{C}\textbf{x}(k)+\textbf{D}\textbf{u}(k)+ \textbf{W}_y \textbf{g}_y\left(\textbf{V}_y^{\text{T}} \left[ \begin{matrix} \textbf{x} \\ \textbf{u} \end{matrix} \right]\right) \label{e:PNLSS2_b},
    \end{empheq}
\end{subequations}
with $\textbf{W}_y \in \mathbb{R}^{p \times r_y}$, $p$ being the size of the output vector, $r_y$ the number of branches in $\textbf{g}_y$, and $\textbf{V}_y \in \mathbb{R}^{(n+m) \times r_y}$.

Recall that the objective of this work is to retrieve a nonlinear representation where $r_x$ and $r_y$ are as small as possible. In the most extreme case $r_x=1$ and $r_y=1$.

Assuming an equivalency\footnote{Conditions on the existence of such an equivalency are stipulated in Section \ref{ss:remarks}.} of the form of Eq.\eqref{e:dec1}, the individual terms of the decoupled formulation can be accessed from the Jacobian information of the coupled function. In that case diagonalising the Jacobian matrix effectively returns both linear transformation matrices $\textbf{V}_x$ and $\textbf{W}_x$. The following relationships can be written for the Jacobian matrix corresponding to a single time sample $k$ (or operating point), 
%
%
%
\begin{equation}
\label{e:Jac1}
\textbf{J}_x(k) = \left[ \begin{matrix} \frac{\partial f_{x_1}(k)}{\partial x_1} \quad \cdots \quad  \frac{\partial f_{x_1}(k)}{\partial x_n}, \quad \frac{\partial f_{x_1}(k)}{\partial u_1}  \quad \cdots \quad \frac{\partial f_{x_1}(k)}{\partial u_m} \\ 
\vdots \quad \quad \ddots  \quad \quad \vdots \quad \  \ \quad \vdots  \quad \quad \ddots  \quad \quad \vdots\\
\frac{\partial f_{x_n}(k)}{\partial x_1} \quad \cdots \quad  \frac{\partial f_{x_n}(k)}{\partial x_n}, \quad \frac{\partial f_{x_n}(k)}{\partial u_1}  \quad \cdots \quad \frac{\partial f_{x_n}(k)}{\partial u_m}
\end{matrix} \right],
\end{equation}%
and \rev{from the decoupled form it follows that}
\begin{equation}
\label{e:Jac2}
\textbf{J}_x(k) = \textbf{W}_x  \operatorname{diag} \left( g'_{x_i}\left( \textbf{v}^{\text{T}}_{x_i} \left[ \begin{matrix} \textbf{x}(k) \\ \textbf{u}(k) \end{matrix}\right]\right) \right) \textbf{V}_x^{\text{T}}.
\end{equation}
Introducing $z_{x_i}(k) = \textbf{v}^{\text{T}}_{x_i} \left[ \begin{matrix} \textbf{x}(k) \\ \textbf{u}(k) \end{matrix}\right]$, the diagonal elements are evaluations of the derivatives of the univariate functions with respect to their arguments $g'_{x_i}(z_{x_i}(k)) = \frac{dg_{x_i}(z_{x_i}(k))}{dz_{x_i}(k)}$. 

In order to parameterise the functions $g'_{x_i}$ one requires a sequence of evaluations for $k=1, \dots, N$. This involves a two step procedure:
\begin{enumerate}
\item Stack the Jacobian matrices, evaluated in $N$ operating points, into a three-way tensor,
\begin{equation}
\label{e:3way_tensor}
 \begin{review} \mathcal{J}_{x_{:,:,k}}\coloneqq \textbf{J}_x(k) \in \mathbb{R}^{n \times (n+m)}. \end{review}
 \end{equation}
\revv{The number of operating points $N$ is not necessarily equal to the data record length. In fact, the required number will depend on whether an exact or an approximate decoupling (see Section \ref{ss:remarks}) is computed. For exact decoupling, following from exact tensor decomposition, $N$ is given by the degree of the coupled function, e.g.\ for a coupled polynomial with monomial terms of degree 2, $N = 3$ is sufficient. 

In the case of an approximate decoupling, following from an approximate tensor decomposition (e.g.\ a regularised CP-decomposition, Section \ref{ss:remarks}), the function approximation will be conditioned on the region which is span by the selected operating points. Therefore good coverage of the operating regime of the function should be ensured. This can be done in two ways, either the operating points are sampled from the training data, or the points are drawn from a distribution which was derived from the training data. In \cite{fakhrizadeh2018,fakhrizadeh2018Phd} it was suggested to draw the operating points $\begin{bmatrix} \mathbf{x}(k) \\ \mathbf{u}(k) \end{bmatrix}$ from normal distributions on \textbf{x} and \textbf{u} with mean and variances corresponding to the training data. The number of points to select then becomes a hyper-parameter which can be explored. In this work $N=1000$ was used by default.}

The major advantage of starting the decoupling from the Jacobian is that a third-order tensor can be used, irrespective of the degree of the polynomials.
\item Perform a simultaneous diagonalisation of all $\mathrm{\mathbf{J}}_x(k)$ \rev{in order to match both expressions of the Jacobian \eqref{e:Jac1} and \eqref{e:Jac2}}.
\end{enumerate}
The latter can be computed using the canonical polyadic decomposition (CP decomposition), implemented in the Tensorlab toolbox \cite{tensorlab} in MATLAB. As such $\mathcal{J}_x$ is decomposed into a sum of rank-one terms,
\begin{equation}
\label{e:rank1}
\mathcal{J}_x = \sum_{i=1}^{r_x} \textbf{w}_{x_i} \circ \textbf{v}_{x_i} \circ \textbf{h}_{x_i},
\end{equation}
where $\circ$ denotes the outer product, \revv{and $\textbf{w}_{x_i}$, $\textbf{v}_{x_i}$, and $\textbf{h}_{x_i}$ are the $i$-th column of the matrices $\textbf{W}_x$, $\textbf{V}_x$, and $\textbf{H}_x$ respectively. Note that the columns in $\mathbf{H}_x \in \mathbb{R}^{N \times r_x}$ are the derivatives of the univariate functions $g_{x_i}$ with respect to their arguments and evaluated in the $N$ operating points.}. As shorthand notation also $\mathcal{J}_x = \llbracket \textbf{W}_x,\textbf{V}_x,\textbf{H}_x \rrbracket$ is used. The minimum number $r_x$ for which the equality in Eq.~\eqref{e:rank1} holds \begin{review} will be\end{review} called the rank of the tensor $\mathcal{J}_x$.

The CP-decomposition immediately returns the matrices $\textbf{W}_x$ and $\textbf{V}_x$. The matrix $\textbf{H}_x$, is however a nonparametric representation of the functions $h_{x_i} = g_{x_i}'(\textbf{z}_{x_i})$ (see Section \ref{ss:avoiding_pr} for a parametrised alternative).  The parametrised functions can be obtained using an ordinary least squares regression, typically a polynomial basis is used, \begin{review} which in turn provides the coefficients of the functions $\textbf{g}_{x}$. \end{review} 

In complete analogy $\textbf{f}_y(\textbf{x},\textbf{u})$ is decomposed into $\textbf{W}_y$, $\textbf{V}_y$ and a set of $r_y$ univariate branches $g_{y_i}(\textbf{z}_{y_i})$.

%

\subsection{Remarks on tensor based polynomial decoupling}
\label{ss:remarks}

In the above it was assumed that, first of all, an equivalency of the form of Eq.~\ref{e:dec1} exists and, secondly, that the decoupled representation could be retrieved via a tensor decomposition. This is however not self-evident and calls for a number of remarks.

We will show that a crucial role is played by the parameter $r$, which is the number of branches used in the decoupled representation, \revv{see e.g.\ Fig.~\ref{f:decoupling}}. It will turn out that whether the decoupling is feasible depends on a set of conditions on $r$ that need to be satisfied simultaneously. The conditions can be subdivided into (a) being inherently related to the coupled function and (b) being related to tensor properties and their decompositions.
\begin{enumerate}[(a)]
\item Given any coupled polynomial function $\textbf{f}(\textbf{x})$, there exists a minimum value of $r$ for which $\textbf{f}(\textbf{x}) =\textbf{W} \textbf{g}(\textbf{z})$ is an exact representation. This can be understood from the fact that all cross-term monomials of the coupled function can be reconstructed from powers of linear combinations of the inputs, e.g.\ consider the function $f(x_1,x_2) = x_1^2x_2$. An exact decoupled representation is given by
\begin{equation}
\label{e:exact_dec}
x_1^2x_2 = \overset{\textbf{W}}{ \left[ \rfrac{1}{6} \quad \rfrac{1}{6} \quad \rfrac{-1}{3} \right]} \overset{\textbf{g}}{\left[ \begin{matrix} z_1^3 \\ z_2^3 \\ z_3^3 \end{matrix} \right]}, \quad \textbf{z} = \overset{\textbf{V}^{\text{T}}}{\left[ \begin{matrix} 1 	\quad 1\\ -1 \quad 1 \\ 0 \quad 1 \end{matrix} \right]}\overset{\textbf{x}}{\left[ \begin{matrix} x_1 \\ x_2 \end{matrix} \right]}.
 \end{equation}
 Here the cross-term is decoupled into 3 univariate branches. Notice that the minimum number of required branches, $r$, grows both with the total degree and with the number of variables used. When $r$ is chosen larger than or equal to this lower bound, it is said that an \emph{exact decoupling} exists.
\item
\begin{enumerate}[1.]
\item When decomposing the Jacobian tensor into rank-one terms (see Eq.~\eqref{e:rank1}), the number of terms, $r$, will define whether the tensor decomposition is exact. A decomposition is considered to be exact when
\begin{equation}
\label{e:forb}
\norm{\mathcal{J} - \llbracket \textbf{W},\textbf{V},\textbf{H} \rrbracket}^2_F=0,
\end{equation}
where subscript $F$ denotes the Frobenius norm. The lowest value of $r$ for which Eq~\eqref{e:forb} holds is defined as the tensor rank. Hence, selecting $r < \text{rank}~\mathcal{J}$ will result in an \emph{approximate decomposition}. An important property of tensors is that it is possible that the $\text{rank}~\mathcal{J}>\text{max}(n_I,n_O,N)$, with $n_I$ the number of inputs and $n_O$ the number of outputs to $\textbf{f}(\textbf{x})$ such that $\mathcal{J}_x \in \mathbb{R}^{n_I \times n_O \times N}$, as opposed to matrix rank, which is limited by the smallest dimension. 

\item Additional to being exact, one also requires the tensor decomposition to be unique \footnote{Only `essential' uniqueness exists, i.e.\ up to a permutation and scaling of the columns of \textbf{V}, \textbf{W} and \textbf{H}.}. Should it not be unique, a simultaneous diagonalisation of the Jacobian matrices would not necessarily return correlated, \rev{i.e.\ polynomial}, evaluations of $g'_i$ along the diagonal (Eq.~\eqref{e:Jac2}).

A sufficient condition for uniqueness, i.e.\ assuming \emph{exact decomposition}, is known as the Kruskal condition \cite{kruskal1977,kruskal1989}. When the operating points in which the Jacobian is evaluated (Eq.~\eqref{e:3way_tensor}) are chosen as random numbers, and for $N>r$, Kruskal's condition boils down to
\begin{equation}
\label{e:kruskal}
\text{min}(n_I,r) + \text{min}(n_O,r) \ge r+2.
\end{equation}
Notice that \revv{neither} the rank of $\mathcal{J}$, nor the degree of the coupled polynomial appear in this sufficient condition. Even for moderate degrees a number of branches $r$ that violate Kruskal's condition can be required in order to meet condition (a). 
\end{enumerate}
\end{enumerate}

\begin{figure}[h!]
\begin{center}

\setlength{\unitlength}{1cm}
\begin{tikzpicture}[scale=0.55] 

%

\node at (0,0) {\footnotesize{CPD}};
\draw [->,thick] (1,0) -- (3,1);
\node at (5,1) {\footnotesize{not unique}};

\draw [->,thick] (6.8,1) -- (8.8,2);
\draw [->,thick] (6.8,1) -- (8.8,0);

\draw [->,thick] (++1,0) -- (++3,-1);
\node at (++5,-1) {\footnotesize{unique}};

\node at (1.8,3) {\footnotesize{(b.2)}};
\node at (7.6,3) {\footnotesize{(a)}};

\node at (10.5,2) {\footnotesize{not exact}};
\node at (10.5,0) {\footnotesize{exact}};

\draw [->,thick] (6.8,-1) -- (8.8,-1);
\node at (10.5,-1) {\footnotesize{exact}};

\end{tikzpicture}
\caption{Scheme that illustrates in what scenarios an exact decoupling of the multivariate polynomial can be retrieved using the CP decomposition. (b.2) and (a) refer to the conditions that are imposed on the number of branches $r$.}
\label{f:kruskal_scheme_tikz}
\end{center}
\end{figure}
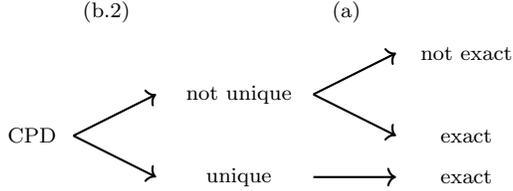

In order to retrieve an \emph{exact decoupling} using tensor decomposition, we hence have that a lower bound on $r$ is provided by \begin{review} condition \end{review}(a) (notice that when (a) is satisfied, so is (b.1)), while a conservative (sufficient condition) upper bound is given by the uniqueness condition of Kruskal (b.2). Whether or not an exact decoupling can be retrieved using the CP decomposition is \begin{review} synthesised \end{review}in Fig.~\ref{f:kruskal_scheme_tikz}.

To illustrate the implications of not fulfilling all conditions we consider again the results obtained for the forced Duffing system (see Section \ref{ss:motivating_ex} and Section \ref{ss:silverbox}). The specifics of the PNLSS model that was created for the forced Duffing system are listed in Table \ref{t:SB}.
\begin{table}[h!]
\caption{Specifics of the PNLSS model of the forced Duffing oscillator}
\label{t:SB}
\begin{center}
\begin{tabular}{| c | c |}
\cline{2-2}
\multicolumn{1}{c|}{} & PNLSS forced Duffing  \\
\hline
$n$ & 2 \\
state nonlinearity & $\textbf{f}_x$, degrees 2, 3\\
output nonlinearity & - \\
rank $\mathcal{J}_x$ & 4 \\
\hline
\hline
$r$ & 4\\
$e_{\text{CPD}}$ & 1.55e-16\\
Kruskal condition & not satisfied\\
\hline
\end{tabular}
\end{center}
\end{table}
It is a second order model with a polynomial nonlinear function of only the state variables in the state equation (the second and third degree monomials are listed in Eq.~\eqref{e:SB1}) and without nonlinearity in the output equation. The relative decomposition error is denoted by $e_{\text{CPD}}$,
\begin{equation}
\label{e:e_CPD}
e_{\text{CPD}} = \frac{\norm{\mathcal{J} - \llbracket \textbf{W},\textbf{V},\textbf{H} \rrbracket}^2_F}{\norm{ \mathcal{J} }^2_F}.
\end{equation}

\begin{figure}
\begin{center}
\includegraphics[width=\textwidth]{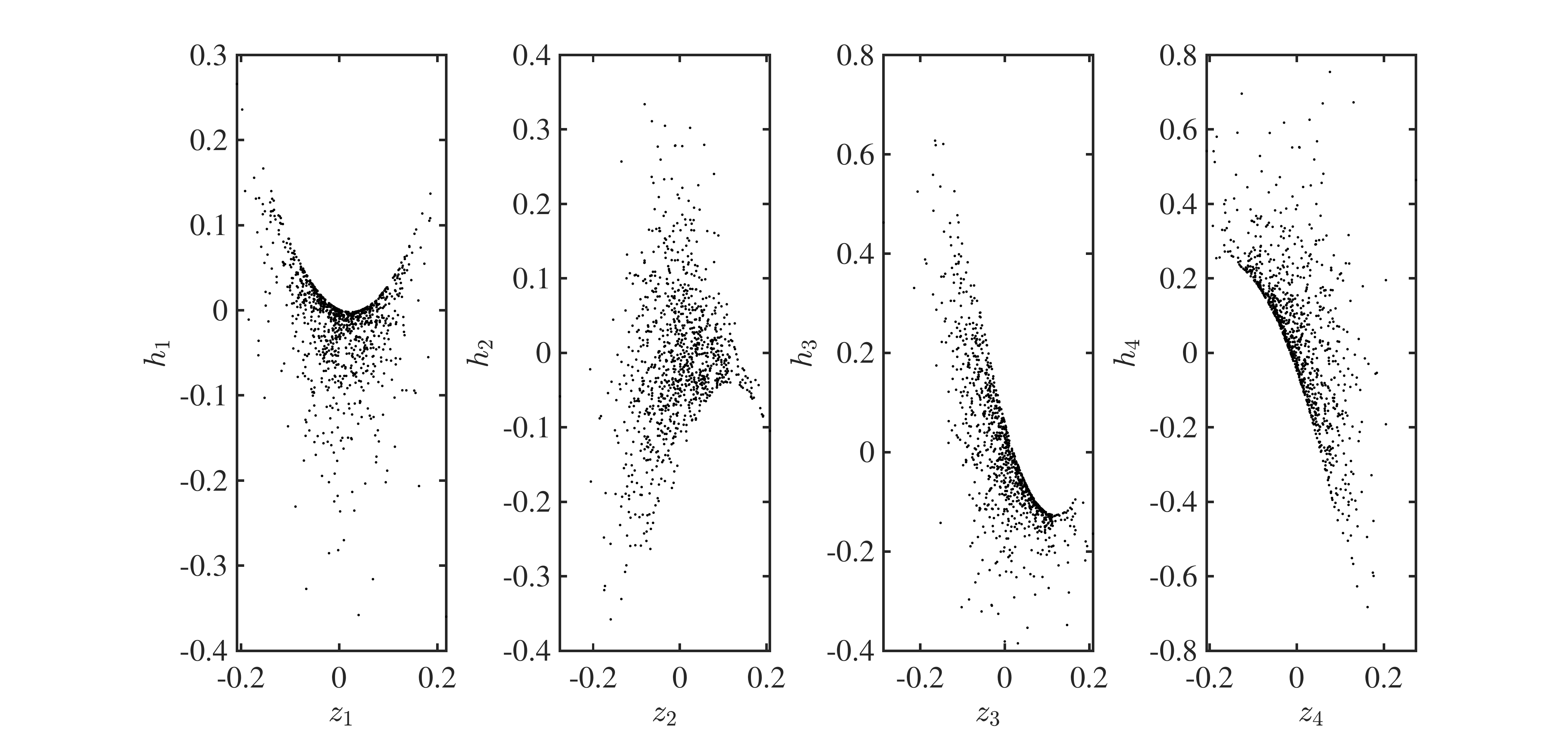}
\caption{Results of the CP-decomposition of $\mathcal{J}_x$ of the PNLSS model obtained for the forced Duffing system. The columns of $\textbf{H}$ (which correspond to the derivatives of the univariate functions \textbf{g}) are plotted as a function of the intermediate variables $z_i$. The large scatter suggests that the decomposition is not unique for $r=4$.}
\label{f:h_cloud_SB}
\end{center}
\end{figure}

If $\textbf{f}_x$ is decomposed into 4 branches, equal to the rank of $\mathcal{J}_x$, a low decomposition error is obtained \begin{review}(by definition \eqref{e:forb})\end{review}. Yet for $r=4$, Kruskal's condition (Eq.~\eqref{e:kruskal}) is not satisfied\revv{: $\text{min}(2,4) + \text{min}(2,4) \ngeqslant 6$}. Hence the uniqueness of the decomposition is not guaranteed. In Fig.~\ref{f:h_cloud_SB} the columns of the obtained \textbf{H}-matrix are shown. The large scatter suggests that the decomposition is not unique \begin{review}and that hence the solution of the CPD does not return nonparametric estimates of $h_i= g'_i$\end{review}. Recall that a parametrisation of the function	$h_i$ is needed in order to retrieve the nonlinear mappings $g_i$. Scatter on $h_i$ will therefore unavoidably result in poor \emph{approximate decoupling}. \revv{It is therefore crucial that Kruskal's condition is checked should one wish to use the classical CPD. Alternatively one can by default resort to one of the options, which do not suffer from the non-uniqueness issue, presented in the next section.} 
%

\subsection{Avoiding large scatter on the sampled function $h$}
\label{ss:avoiding_pr}
 A number of solutions that either reduce or avoid scatter on the function $h$ have been proposed. This section provides an overview.

\begin{itemize}
\item \textbf{Smoothness promoting regularised tensor decomposition}

The CP decomposition relies on an alternating least-squares procedure to return \textbf{V}, \textbf{W} and \textbf{H}. By penalising high finite differences amongst the successive elements in the columns of \textbf{H}, smoothness is promoted. In many cases the gain on the fit of the $h$-functions far outweighs the loss in accuracy of the approximate tensor decomposition. The idea, proposed by Dreesen, can be consulted in \cite{decuyper2019}. \revv{It is suggested to use such a regularised CP-decomposition by default. Should the uniqueness condition nevertheless be satisfied, a search of the hyperparameters will return parameter values equal to zero, hence collapsing the method to the classical CP-decomposition.}

\item \textbf{Parameterised tensor decomposition}

In \cite{hollander2018} it was proposed to use a polynomial constraint on the columns of \textbf{H} while computing the CP decomposition. Doing so bypasses the fitting step, hence avoiding the issue entirely. The price to pay is loss of the monotonic convergence properties of the alternating least-squares optimisation, potentially deteriorating the \begin{review} tensor approximation of Eq.~\eqref{e:e_CPD} ($e_{\text{CPD}}$)\end{review}.

\item \textbf{Structured Data Fusion}

In \cite{dreesen2018} first-order and second-order derivative information are combined into a joint tensor decomposition with partial symmetry. This can be phrased into the Structured Data Fusion framework \cite{sorber2015}. It is expected that imposing additional constraints on the decomposition will ensure that uniqueness conditions are more easily met.
\end{itemize}

\section{Reducing decoupled nonlinear state-space models}
\label{s:reduction}

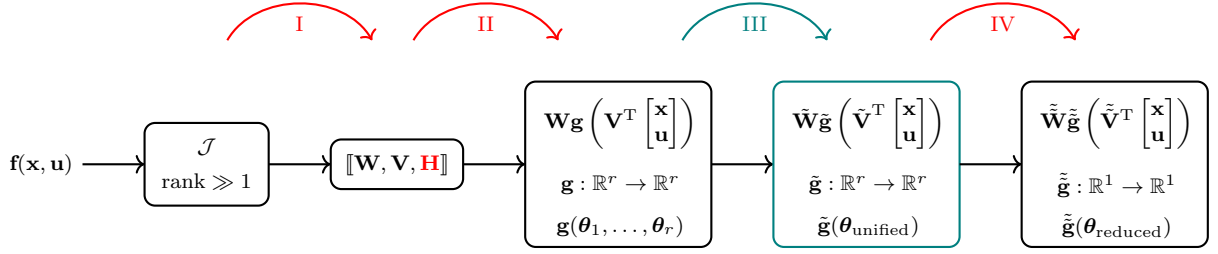
\begin{figure}[h]
\begin{center}

\setlength{\unitlength}{1cm}
\begin{tikzpicture}[scale=0.55] 


%
%
%
%
%
%
%
%


\node at (0,0) {\footnotesize{$\textbf{f}(\textbf{x},\textbf{u})$}};
\draw [->,thick] (1,0) -- (2.5,0);

\node [above] at (4,0) {\footnotesize{$\mathcal{J}$}};
\node [below] at (4,0) {\footnotesize{$\text{rank}\gg1$}};
\draw[thick,rounded corners] (2.5,0) -- (2.5,1) -- (5.5,1) -- (5.5,-1) -- (2.5,-1) -- (2.5,0);
\draw [->,thick] (5.5,0) -- (7,0);

\node at (8.6,0) {\footnotesize{$\llbracket \textbf{W}, \textbf{V}, \color{red} \textbf{H} \color{black} \rrbracket$}};
\draw[thick,rounded corners] (7,0) -- (7,0.6) -- (10.2,0.6) -- (10.2,-0.6) -- (7,-0.6) -- (7,0);
\draw [->,thick] (10.2,0) -- (11.7,0);

\node at (14,1) {\footnotesize{$\textbf{W}\textbf{g}\left( \textbf{V}^{\text{T}} \left[ \begin{matrix} \textbf{x} \\ \textbf{u} \end{matrix} \right] \right)$}};
\node at (14,-0.5) {\footnotesize{$\textbf{g}: \mathbb{R}^r \rightarrow \mathbb{R}^r$}};
\node at (14,-1.5) {\footnotesize{$\textbf{g}(\bm{\theta}_1,\dots,\bm{\theta}_r)$}};
\draw[thick,rounded corners] (11.7,0) -- (11.7,2) -- (16.2,2) -- (16.2,-2) -- (11.7,-2) -- (11.7,0);
\draw [->,thick] (16.2,0) -- (17.7,0);

\node at (20,1) {\footnotesize{$\tilde{\textbf{W}}\tilde{\textbf{g}}\left( \tilde{\textbf{V}}^{\text{T}} \left[ \begin{matrix} \textbf{x} \\ \textbf{u} \end{matrix} \right] \right)$}};
\node at (20,-0.5) {\footnotesize{$\tilde{\textbf{g}}: \mathbb{R}^r \rightarrow \mathbb{R}^r$}};
\node at (20,-1.5) {\footnotesize{$\tilde{\textbf{g}}(\bm{\theta}_{\text{unified}})$}};
\draw[thick,teal,rounded corners] (17.7,0) -- (17.7,2) -- (22.2,2) -- (22.2,-2) -- (17.7,-2) -- (17.7,0);
\draw [->,thick] (22.2,0) -- (23.7,0);

\node at (26,1) {\footnotesize{$\tilde{\tilde{\textbf{W}}}\tilde{\tilde{\textbf{g}}}\left( \tilde{\tilde{\textbf{V}}}^{\text{T}} \left[ \begin{matrix} \textbf{x} \\ \textbf{u} \end{matrix} \right] \right)$}};
\node at (26,-0.5) {\footnotesize{$\tilde{\tilde{\textbf{g}}}: \mathbb{R}^1 \rightarrow \mathbb{R}^1$}};
\node at (26,-1.5) {\footnotesize{$\tilde{\tilde{\textbf{g}}}(\bm{\theta}_{\text{reduced}})$}};
\draw[thick,rounded corners] (23.7,0) -- (23.7,2) -- (28.2,2) -- (28.2,-2) -- (23.7,-2) -- (23.7,0);

\draw[red, thick,->] (4.5,3) to [out=60,in=120] (8,3);
\node [above] at (6.25,3) {\footnotesize{\color{red}{I}}};
\draw[red, thick,->] (9,3) to [out=60,in=120] (12.5,3);
\node [above] at (10.75,3) {\footnotesize{\color{red}{II}}};
\draw[teal, thick,->] (15.5,3) to [out=60,in=120] (19,3);
\node [above] at (17.25,3) {\footnotesize{\color{teal}{III}}};
\draw[red, thick,->] (21.5,3) to [out=60,in=120] (25,3);
\node [above] at (23.25,3) {\footnotesize{\color{red}{IV}}};

\end{tikzpicture}
\caption{Model reduction procedure. Roman numbers indicate the steps: I regularised CP decomposition, II parametrisation of the nonlinear mappings \textbf{g}, III optional step where all branches are unified, IV branch reduction down to $r=1$.}
\label{f:scheme_red}
\end{center}
\end{figure}

Given the large number of degrees of freedom that are provided during black-box identification it is likely that an overly complex model is obtained. One could therefore attempt to simplify or reduce this model in a proceeding step. As a starting point we will use models with decoupled nonlinear parts which have a large number of branches. The decoupling is assumed to have taken into account the remarks of Section \ref{ss:remarks} by any of the solutions mentioned in Section \ref{ss:avoiding_pr}. Two straightforward actions will be discussed: 
\begin{enumerate}
\item Force all branches to take a single functional form. We will refer to this action as \emph{unifying the branches}.
\item Reduce the number of branches.
\end{enumerate}
A schematic overview of the approach is provided in Fig.~\ref{f:scheme_red}. The roman numbers indicate the following steps,
\begin{enumerate}[I]
\item When constructing the Jacobian tensor it is typically of $\text{rank} \gg 1$. To limit $e_{\text{CPD}}$ a decomposition into $r=\text{rank}~\mathcal{J}$ is computed. Step I represents such decomposition where regularisation is used to promote smoothness on the columns of \textbf{H} (Section \ref{ss:avoiding_pr}) \cite{decuyper2019}.
\item $r$ nonlinear mapping functions $\textbf{g}$ are obtained from integrating the parameterised $h$-functions, i.e.\ the columns of \textbf{H} which have been fitted with polynomials (Section \ref{ss:decoupling}).
\item All branches are unified such that $r$ identical branches remain. This is an optional step, to be used only when deemed fit (Section \ref{ss:unifying}). 
\item The number of branches is reduced to $r=1$ (Section \ref{ss:branch_reduction}). 
\end{enumerate}


Each step is followed by optimising all model parameters on a model output error level (Eq.~\eqref{e:VLS}). The model reduction \rev{progresses in successive steps and can be stopped prematurely (i.e.\ with $r>1$) when the model performance falls below a threshold.} The method allows to balance model complexity with accuracy.


\subsection{Unifying the univariate branches}
\label{ss:unifying}

From visual inspection of the branches it may turn out that the nonlinear mappings \textbf{g} are closely related. In such cases a reduction in the number of parameters can be achieved by imposing a single functional form on the branches. \revv{This is achieved in two steps:
\begin{enumerate}
\item Appoint one of the branches to be the unified form and replace all others with the identical function. This will introduce an error at the function level. Optionally, the introduced error may be limited by scaling the input and the output of each branch. This can be done by hand or by using an optimisation algorithm. The result are improved starting values for the full function optimisation of step 2.
\item Consider step 1 to be an initialisation step and minimise the introduced error by solving the optimisation problem of Eq.~\eqref{e:uni}. Solving Eq.~\eqref{e:uni} minimises the output error of the unified function with respect to the original output.
\end{enumerate}
\begin{equation}
\label{e:uni}
\underset{\tilde{\textbf{V}},\tilde{\textbf{W}},\bm{\theta}_{\text{unified}}}{\operatorname{minimise}}~\sum_{k=1}^N\left\| \textbf{q}(k) - \tilde{\textbf{W}}\tilde{\textbf{g}}\left(\tilde{\textbf{V}}^{\text{T}} \left[ \begin{matrix} \textbf{x}(k) \\ \textbf{u}(k) \end{matrix} \right], \bm{\theta}_{\text{unified}}\right)\right\|^2_2,
\end{equation}
here $\bm{\theta}_{\text{unified}}$ are the parameters of $\tilde{\textbf{g}}$ and $\textbf{q}$ is the output of the original function, i.e.\ before unification of the branches. The optimisation is initialised using the original \textbf{V} and \textbf{W}, and the parameters of the selected branch in step 1. An example on a fictitious decoupled function is provided in Table \ref{t:unified_example}.
\begin{table}[h!]
\begin{center}
\caption{\color{black} Example of the unifying step on a fictitious decoupled function.}
\label{t:unified_example}
\begin{tabular}{c | c | c }
\color{black} original decoupled function & \color{black}1.\ unified initialisation & \color{black}2.\ optimised unified function \\
\hline
\color{black}$\begin{matrix} \textbf{V} = \left[\begin{matrix} 0.87~0.35~0.56\\ 0.11~0.24~0.61 \end{matrix} \right] \\ \\ \textbf{W} = \left[ \begin{matrix} 0.60~ 0.52~ 0.69 \\ 0.60~0.01~0.95 \end{matrix} \right] \\ \\ 
\textbf{g} = \left[ \begin{matrix} 0.3z_1^3+0.5z_1^2 \\  -0.28z_2^3-0.48z_2^2 \\ 0.25z_3^3+0.45z_3^2\end{matrix} \right]\end{matrix}$ & $\begin{matrix} \\ \color{black} \textbf{V} = \left[\begin{matrix} 0.87~0.35~0.56\\ 0.11~0.24~0.61 \end{matrix} \right]\\ \\  \color{black} \textbf{W} = \left[ \begin{matrix} 0.60~ 0.52~ 0.69 \\ 0.60~0.01~0.95 \end{matrix} \right] \\ \\
\color{black} \tilde{\textbf{g}} = \left[ \begin{matrix} 0.3z_1^3+0.5z_1^2 \\  0.3z_2^3+0.5z_2^2 \\ 0.3z_3^3+0.5z_3^2\end{matrix} \right] \\ \\  \color{black} e_{f1} = 0.21 \\  \color{black}e_{f2} =  0.11 \end{matrix}$ & $\begin{matrix} \\ \color{black} \tilde{\textbf{V}} = \left[ \begin{matrix} 0.66~ 0.18~0.52\\ 0.28~0.08~0.88 \end{matrix} \right]
 \\ \\\color{black} \tilde{\textbf{W}} = \left[ \begin{matrix} 0.79~0.33~0.68 \\ 0.46~0.78~0.24 \end{matrix} \right] \\ \\
 \color{black} \tilde{\textbf{g}} = \left[\begin{matrix} 0.26z_1^3+0.44z_1^2\\0.26z_2^3+0.44z_2^2\\0.26z_3^3+0.44z_3^2\end{matrix} \right] \\ \\ \color{black} e_{f1} = 0.003 \\ \color{black} e_{f2} = 0.003  \end{matrix}$ \\
\end{tabular}
\end{center}
\end{table}

During step one, in the example, the first component of \textbf{g} was appointed as the unified functional form. This introduced a function error measured using a relative root-mean-squared metric $e_f = \frac{\text{rms}(\textbf{q} - \tilde{\textbf{q}})}{\text{rms}(\textbf{q})}$, with $\tilde{\textbf{q}}$ representing the output of the altered function. Step two reduces the relative error to below 1\%. The original decoupled function and the ultimately obtained unified function are additionally depicted in Fig.~\ref{f:unified_example}.
\begin{figure}
\begin{center}
\begin{subfigure}[b]{0.45\textwidth}
\begin{center}
\includegraphics[width=\textwidth]{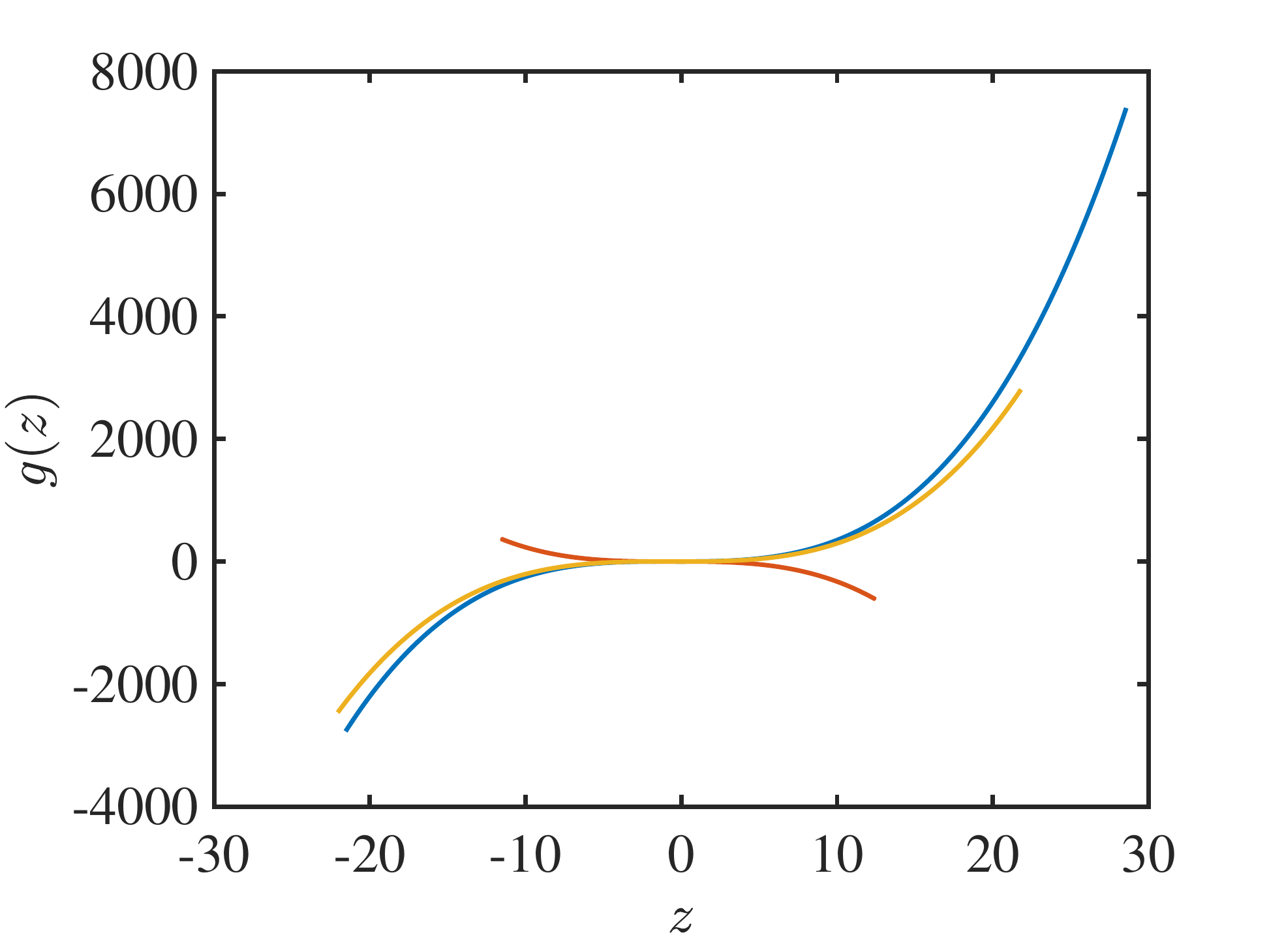}
\caption{}
\end{center}
\end{subfigure}
\begin{subfigure}[b]{0.45\textwidth}
\begin{center}
\includegraphics[width=\textwidth]{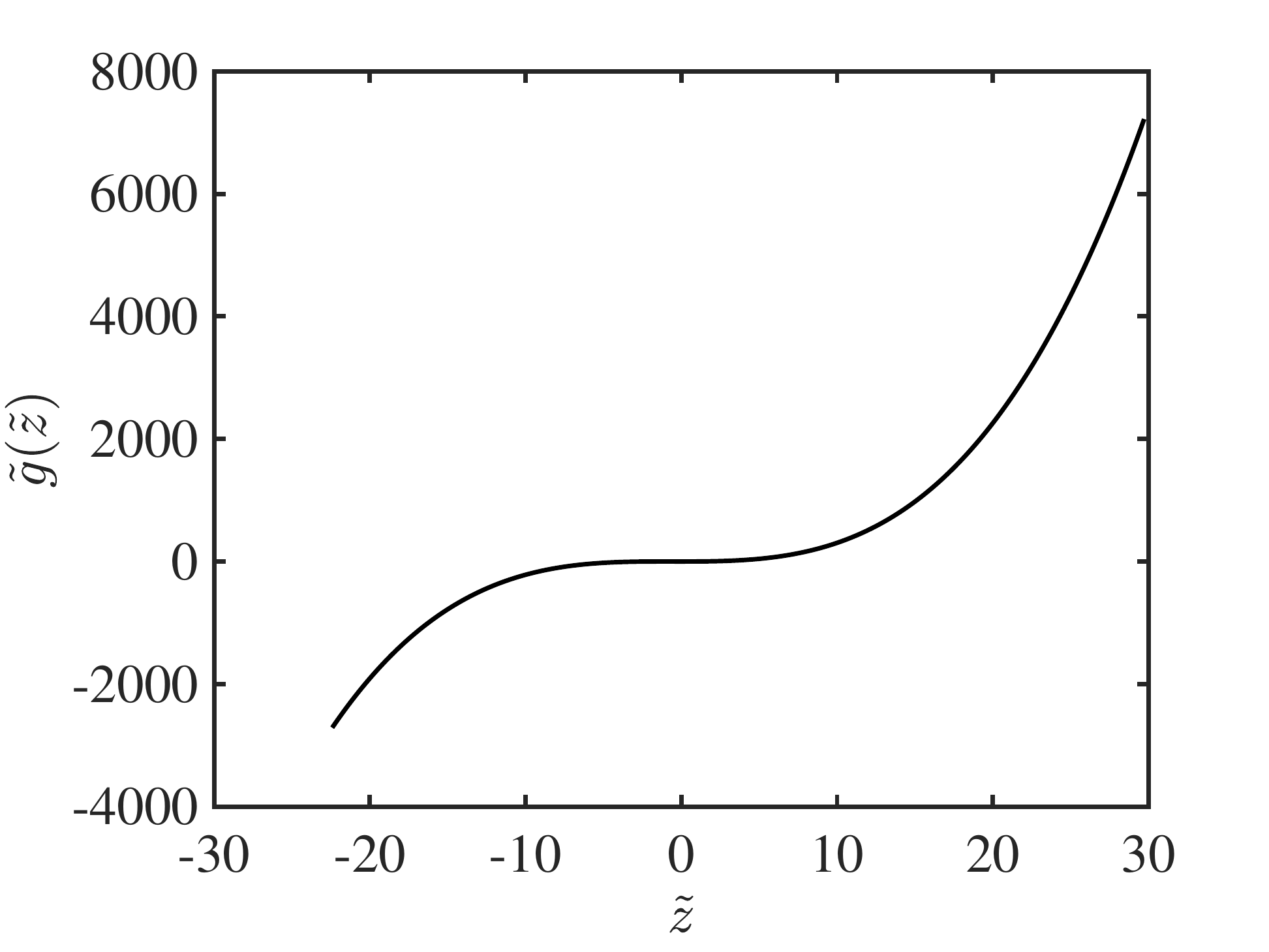}
\caption{}
\end{center}
\end{subfigure}
\caption{\color{black}Functions corresponding to Table \ref{t:unified_example}. (a) Plot of the three resembling branch functions. (b) Plot of the ultimately (after optimisation) obtained unified branch.}
\label{f:unified_example}
\end{center}
\end{figure}}

A low output error of the unified function is necessary in order to serve as a good initialisation, for further optimisation (see Section \ref{ss:fine_tuning}). A deviation of \revv{output} of the unified function from the original \revv{output} can moreover trigger unstable response \begin{review}when plugged back into the state-space model\end{review}. 

\begin{figure}[h]
\begin{center}

\setlength{\unitlength}{1cm}
\begin{tikzpicture}[scale=0.55] 


%
%
%
%
%
%
%
%


\node at (-0.5,4.5) {\scriptsize{$x_1$}};
\node at (-0.5,3) {\scriptsize{$x_n$}};

\node at (-0.5,2) {\scriptsize{$u_1$}};
\node at (-0.5,0.5) {\scriptsize{$u_m$}};

\draw [->,thick] (0,4.5) -- (1,4.5);
\draw[dotted,thick] (0.5,3.55) -- (0.5,3.95);
\draw [->,thick] (0,3) -- (1,3);
\draw [->,thick] (0,2) -- (1,2);
\draw[dotted,thick] (0.5,1.05) -- (0.5,1.45);
\draw [->,thick] (0,0.5) -- (1,0.5);

\draw [thick] (1,0) rectangle(2.5,5);
\node at (1.75,2.5) {$\textbf{V}_x^{\text{T}}$};

\draw [->,thick] (2.5,4) -- (4,4);
\draw[dotted,thick] (5,2.4) -- (5,2.8);
\draw [->,thick] (2.5,1) -- (4,1);


\draw[thick,rounded corners, fill=lightgray] (4,4) -- (4,4.475) -- (6,4.475) -- (6,3.525) -- (4,3.525) -- (4,4);
\node at (5,4) {\scriptsize{$g_{x_1}$}\scriptsize{$(z_1)$}};

\draw[thick,rounded corners, fill=lightgray] (4,1) -- (4,1.475) -- (6,1.475) -- (6,0.525) -- (4,0.525) -- (4,1);
\node at (5,1) {\scriptsize{$g_{x_r}$}\scriptsize{$(z_r)$}};

\draw [->,thick] (6,4) -- (7.5,4);
\draw [->,thick] (6,1) -- (7.5,1);

\draw [thick] (7.5,0) rectangle(9,5);
\node at (8.25,2.5) {$\textbf{W}_x$};

\draw [->,thick] (9,4) -- (10,4);
\draw[dotted,thick] (9.5,2.4) -- (9.5,2.8);
\draw [->,thick] (9,1) -- (10,1);

\node at (10.6,4) {\scriptsize{$q_1$}};
\node at (10.6,1) {\scriptsize{$q_n$}};

\draw[<->, ultra thick] (12,2.5) -- (13.5,2.5);


\node at (14.5,4.5) {\scriptsize{$x_1$}};
\node at (14.5,3) {\scriptsize{$x_n$}};

\node at (14.5,2) {\scriptsize{$u_1$}};
\node at (14.5,0.5) {\scriptsize{$u_m$}};

\draw [->,thick] (15,4.5) -- (16,4.5);
\draw[dotted,thick] (15.5,3.55) -- (15.5,3.95);
\draw [->,thick] (15,3) -- (16,3);
\draw [->,thick] (15,2) -- (16,2);
\draw[dotted,thick] (15.5,1.05) -- (15.5,1.45);
\draw [->,thick] (15,0.5) -- (16,0.5);

\draw [thick] (16,0) rectangle(17.5,5);
\node at (16.75,2.5) {$\tilde{\textbf{V}}_x^{\text{T}}$};

\draw [->,thick] (17.5,4) -- (19,4);
\draw[dotted,thick] (20,2.4) -- (20,2.8);
\draw [->,thick] (17.5,1) -- (19,1);


\draw[thick,rounded corners, fill=lightgray] (19,4) -- (19,4.475) -- (21,4.475) -- (21,3.525) -- (19,3.525) -- (19,4);
\node at (20,4) {\scriptsize{$\tilde{g}_{\text{uni}}$}\scriptsize{$(\tilde{z}_1)$}};

\draw[thick,rounded corners, fill=lightgray] (19,1) -- (19,1.475) -- (21,1.475) -- (21,0.525) -- (19,0.525) -- (19,1);
\node at (20,1) {\scriptsize{$\tilde{g}_{\text{uni}}$}\scriptsize{$(\tilde{z}_r)$}};

\draw [->,thick] (21,4) -- (22.5,4);
\draw [->,thick] (21,1) -- (22.5,1);

\draw [thick] (22.5,0) rectangle(24,5);
\node at (23.25,2.5) {$\tilde{\textbf{W}}_x$};

\draw [->,thick] (24,4) -- (25,4);
\draw[dotted,thick] (24.5,2.4) -- (24.5,2.8);
\draw [->,thick] (24,1) -- (25,1);

\node at (25.6,4) {\scriptsize{$\tilde{q}_1$}};
\node at (25.6,1) {\scriptsize{$\tilde{q}_n$}};

\end{tikzpicture}
\caption{Left: a decoupled representation of a state equation nonlinearity, $\textbf{f}_x$. Right: a reduction of the decoupled function where the nonlinear mappings \color{black}($\tilde{g}$) are unified, i.e.\ they are identical.}
\label{f:unified1}
\end{center}
\end{figure}
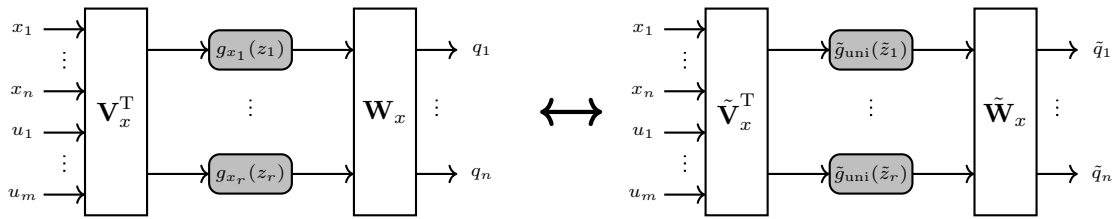

A graphical illustration of a \emph{unified decoupled} nonlinear function is provided in Fig.~\ref{f:unified1}.

Notice that having identical branches results in a function which resembles a neural network with one hidden layer and a polynomial activation function.


\subsection{Reducing the number of branches}
\label{ss:branch_reduction}

An intuitive way of reducing the model is by reducing the number of branches. The number of branches will generally exceed the minimal number required in order to attain an equivalent model accuracy. 

In order to remove a branch we will exploit the fact that, first of all, not all branches contribute equally to the function output and secondly, that the branches are not necessarily linearly independent \revv{from one another}. Hence a linear combination of the remaining branches can be sought so to minimise the impact of removing a branch. The latter corresponds to an update of the \textbf{W}-matrix. \revv{This is clarified in what follows, let $\textbf{q}$ be the output of the original function, i.e.\ before removing any of the branches,}
\begin{equation}
\color{black} \textbf{q} = \textbf{W}\textbf{g}\left(\textbf{V}^{\text{T}} \left[\begin{matrix} \textbf{x} \\ \textbf{u} \end{matrix} \right] \right),
\end{equation}
\revv{with $\textbf{W} \in \mathbb{R}^{n_o \times r}$, $\textbf{g}: \mathbb{R}^r \rightarrow \mathbb{R}^r$, and $\textbf{V} \in \mathbb{R}^{n_I \times r}$, and let $\tilde{\tilde{\textbf{q}}}_0$ be the output of the reduced function where one of the branches has been removed,}
\begin{equation}
\color{black} \tilde{\tilde{\textbf{q}}}_0 = \tilde{\tilde{\textbf{W}}}_0\tilde{\tilde{\textbf{g}}}\left(\tilde{\tilde{\textbf{V}}}^{\text{T}} \left[\begin{matrix} \textbf{x} \\ \textbf{u} \end{matrix} \right] \right),
\end{equation}
\revv{hence $\tilde{\tilde{\textbf{W}}}_0 \in \mathbb{R}^{n_o \times (r-1)}$, the subscript 0 is added to denote that this is a plain reduction of \textbf{W}, removing one of its columns, which will serve as an initialisation for a next step. Furthermore $\tilde{\tilde{\textbf{g}}}: \mathbb{R}^{r-1} \rightarrow \mathbb{R}^{r-1}$, and $\tilde{\tilde{\textbf{V}}} \in \mathbb{R}^{n_I \times (r-1)}$. We can then minimise $\left\| \textbf{q} - \tilde{\tilde{\textbf{q}}}_0 \right\|_2^2$ by solving a linear least-squares problem,}
\begin{equation}
\color{black} \textbf{q} - \tilde{\tilde{\textbf{q}}}_0 = \bm{\Delta}_{\textbf{W}} \tilde{\tilde{\textbf{g}}}\left(\tilde{\tilde{\textbf{V}}}^{\text{T}} \left[\begin{matrix} \textbf{x} \\ \textbf{u} \end{matrix} \right] \right),
\end{equation}
\revv{where we solve for $\bm{\Delta}_{\textbf{W}}$. This corresponds to reconstructing the output of the branch which was removed from a linear combination of the remaining branches. The parameters of $\tilde{\tilde{\textbf{W}}}$ are then found from the update,}
\begin{equation}
\color{black}\tilde{\tilde{\textbf{W}}} = \tilde{\tilde{\textbf{W}}}_0 + \bm{\Delta}_{\textbf{W}},
\end{equation}
\revv{and the output of the reduced function is given by,}
\begin{equation}
\color{black} \tilde{\tilde{\textbf{q}}} = \tilde{\tilde{\textbf{W}}}\tilde{\tilde{\textbf{g}}}\left(\tilde{\tilde{\textbf{V}}}^{\text{T}} \left[\begin{matrix} \textbf{x} \\ \textbf{u} \end{matrix} \right] \right).
\end{equation}
\revv{To decide which branch to remove, the ultimate impact of the action is assessed for all the branches, i.e.\ look for the branch which after reduction and update of $\tilde{\tilde{\textbf{W}}}_0$ results in the lowest relative output error $e_f = \frac{\text{rms}(\textbf{q} - \tilde{\tilde{\textbf{q}}})}{\text{rms}(\text{q})}$. This will favour the removal of two types of branches, those which contribute less to the output of the function and those which are linearly dependent on others.

Additionally one could add a nonlinear optimisation step (at the function level), similar to Eq.~\eqref{e:uni},
\begin{equation}
\label{e:red}
\color{black} \underset{\tilde{\tilde{\textbf{V}}},\tilde{\tilde{\textbf{W}}},\bm{\theta}_{\text{reduced}}}{\operatorname{minimise}}~\sum_{k=1}^N\left\| \textbf{q}(k) - \tilde{\tilde{\textbf{W}}}\tilde{\tilde{\textbf{g}}}\left(\tilde{\tilde{\textbf{V}}}^{\text{T}} \left[ \begin{matrix} \textbf{x}(k) \\ \textbf{u}(k) \end{matrix} \right], \bm{\theta}_{\text{reduced}}\right)\right\|^2_2.
\end{equation}}

\begin{figure}[h]
\begin{center}

\setlength{\unitlength}{1cm}
\begin{tikzpicture}[scale=0.55] 
\node at (-0.5,4.5) {\scriptsize{$x_1$}};
\node at (-0.5,3) {\scriptsize{$x_n$}};

\node at (-0.5,2) {\scriptsize{$u_1$}};
\node at (-0.5,0.5) {\scriptsize{$u_m$}};

\draw [->,thick] (0,4.5) -- (1,4.5);
\draw[dotted,thick] (0.5,3.55) -- (0.5,3.95);
\draw [->,thick] (0,3) -- (1,3);
\draw [->,thick] (0,2) -- (1,2);
\draw[dotted,thick] (0.5,1.05) -- (0.5,1.45);
\draw [->,thick] (0,0.5) -- (1,0.5);

\draw [thick] (1,0) rectangle(2.5,5);
\node at (1.75,2.5) {$\textbf{V}^{\text{T}}$};

\draw [->,thick] (2.5,4) -- (4,4);
\draw[dotted,thick] (5,2.4) -- (5,2.8);
\draw [->,thick] (2.5,1) -- (4,1);


\draw[thick,rounded corners, fill=lightgray] (4,4) -- (4,4.475) -- (6,4.475) -- (6,3.525) -- (4,3.525) -- (4,4);
\node at (5,4) {\scriptsize{$g_1$}\scriptsize{$(z_1)$}};

\draw[thick,rounded corners, fill=lightgray] (4,1) -- (4,1.475) -- (6,1.475) -- (6,0.525) -- (4,0.525) -- (4,1);
\node at (5,1) {\scriptsize{$g_r$}\scriptsize{$(z_r)$}};

\draw [->,thick] (6,4) -- (7.5,4);
\draw [->,thick] (6,1) -- (7.5,1);

\draw [thick] (7.5,0) rectangle(9,5);
\node at (8.25,2.5) {$\textbf{W}$};

\draw [->,thick] (9,4) -- (10,4);
\draw[dotted,thick] (9.5,2.4) -- (9.5,2.8);
\draw [->,thick] (9,1) -- (10,1);

\node at (10.6,4) {\scriptsize{$q_1$}};
\node at (10.6,1) {\scriptsize{$q_n$}};

\draw[<->, ultra thick] (12,2.5) -- (13.5,2.5);


\node at (14.5,4.5) {\scriptsize{$x_1$}};
\node at (14.5,3) {\scriptsize{$x_n$}};

\node at (14.5,2) {\scriptsize{$u_1$}};
\node at (14.5,0.5) {\scriptsize{$u_m$}};

\draw [->,thick] (15,4.5) -- (16,4.5);
\draw[dotted,thick] (15.5,3.55) -- (15.5,3.95);
\draw [->,thick] (15,3) -- (16,3);
\draw [->,thick] (15,2) -- (16,2);
\draw[dotted,thick] (15.5,1.05) -- (15.5,1.45);
\draw [->,thick] (15,0.5) -- (16,0.5);

\draw [thick] (16,0) rectangle(17.5,5);
\node at (16.75,2.5) {$\tilde{\tilde{\textbf{v}}}^{\text{T}}$};

\draw [->,thick] (17.5,2.5) -- (19,2.5);


\draw[thick,rounded corners, fill=lightgray] (19,2.5) -- (19,2.975) -- (21,2.975) -- (21,2.025) -- (19,2.025) -- (19,2.5);
\node at (20,2.5) {\scriptsize{$\tilde{\tilde{g}}$}\scriptsize{$(\tilde{\tilde{z}})$}};


\draw [->,thick] (21,2.5) -- (22.5,2.5);

\draw [thick] (22.5,0) rectangle(24,5);
\node at (23.25,2.5) {$\tilde{\tilde{\textbf{w}}}$};

\draw [->,thick] (24,4) -- (25,4);
\draw[dotted,thick] (24.5,2.4) -- (24.5,2.8);
\draw [->,thick] (24,1) -- (25,1);

\node at (25.6,4) {\scriptsize{$\tilde{\tilde{q}}_1$}};
\node at (25.6,1) {\scriptsize{$\tilde{\tilde{q}}_n$}};

\end{tikzpicture}
\caption{Left: a decoupled representation of a state equation nonlinearity, $\textbf{f}_x$. Right: a reduction down to a single univariate branch where the linear transformation matrices have become vectors.}
\label{f:reduced}
\end{center}
\end{figure}
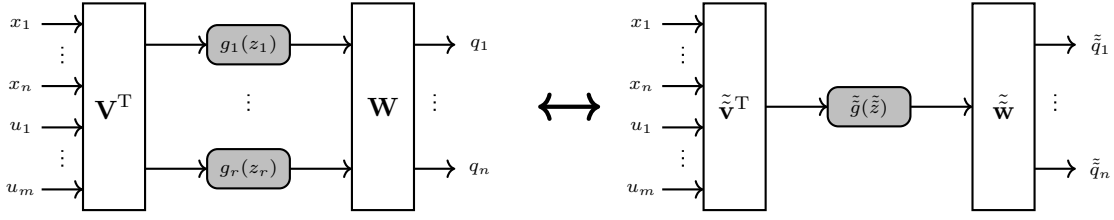

The reduction proceeds in alternating steps, i.e.\ after removing a branch, the reduced function is implemented back into the state-space model and an overal optimisation on a model-output-error level is executed (Section \ref{ss:fine_tuning}). Hence the model is gradually cast into a reduced form. An illustration of a \begin{review}single-branch \end{review}model is provided in Fig.~\ref{f:reduced}.


%

\revv{An example of a single reduction step (removing one branch) on a fictitious decoupled function is provided in Table \ref{t:reduced_example}. In this example, updating the $\tilde{\tilde{\textbf{W}}}$ matrix results in a good approximation of the original function output. Therefore no additional nonlinear optimisation step (Eq.\eqref{e:red}) was computed.}
\begin{table}[h!]
\begin{center}
\caption{\color{black} Example of the branch reduction step on a fictitious decoupled function.}
\label{t:reduced_example}
\begin{tabular}{c | c | c }
\color{black} original decoupled function & \color{black}1.\ reduced initialisation & \color{black}2.\ updated reduced function \\
\hline
\color{black}$\begin{matrix} \textbf{V} = \left[\begin{matrix} 0.87~0.35~0.56\\ 0.11~0.24~0.61 \end{matrix} \right] \\ \\ \textbf{W} = \left[ \begin{matrix} 0.60~ 0.52~ 0.69 \\ 0.60~0.01~0.95 \end{matrix} \right] \\ \\ 
\textbf{g} = \left[ \begin{matrix} 0.3z_1^3+0.5z_1^2 \\  -0.28z_2^3-0.48z_2^2 \\ 0.25z_3^3+0.45z_3^2\end{matrix} \right]\end{matrix}$ & \color{black} $ \begin{matrix} \\ \tilde{\tilde{\textbf{V}}} = \left[\begin{matrix} 0.87~0.56\\ 0.11~0.61 \end{matrix} \right]\\ \\  \color{black} \tilde{\tilde{\textbf{W}}}_0 = \left[ \begin{matrix} 0.60~ 0.69 \\ 0.60~0.95 \end{matrix} \right] \\ \\
\color{black} \tilde{\tilde{\textbf{g}}} = \left[ \begin{matrix} 0.3z_1^3+0.5z_1^2  \\0.25z_3^3+0.45z_3^2\end{matrix} \right] \\ \\  \color{black} e_{f1} = 0.06 \\  \color{black}e_{f2} =  0.0007 \end{matrix}$ & $\begin{matrix} \\ \color{black} \tilde{\tilde{\textbf{V}}} = \left[ \begin{matrix} 0.87~0.56\\ 0.11~0.61 \end{matrix} \right]
 \\ \\\color{black} \tilde{\tilde{\textbf{W}}} = \left[ \begin{matrix} 0.56~0.63 \\ 0.60~0.95\end{matrix} \right] \\ \\
 \color{black} \tilde{\tilde{\textbf{g}}} = \left[\begin{matrix} 0.3z_1^3+0.5z_1^2\\0.25z_3^3+0.45z_3^2\end{matrix} \right] \\ \\ \color{black} e_{f1} = 0.008 \\ \color{black} e_{f2} = 0.00009  \end{matrix}$ \\
\end{tabular}
\end{center}
\end{table}

%

\subsection{Fine-tuning of the decoupled nonlinear state-space model}
\label{ss:fine_tuning}

Note that the result of the decoupling and/or reduction procedure is plugged back into the state-space model. This model then serves as an initialisation of which all parameters ($\bm{\theta}_{ss}$) undergo further optimisation on a model-output-error level. The latter requires nonlinear optimisation of the least-squares cost functions ($\textbf{V}_{\text{LS}}$ given by Eq.~\eqref{e:VLS}) and is solved using a Levenberg-Marquardt algorithm. The vector of parameters to be tuned, in case of a decoupled model, is given by
%
%
\begin{equation}
\centering
\begin{multlined}
\begin{aligned}
\bm{\theta}_{ss} = [\text{vec}(\textbf{A});\text{vec}(\textbf{B});\text{vec}(\textbf{C});\text{vec}(\textbf{D});\text{vec}(\textbf{W}_x);\\
\text{vec}(\textbf{W}_y);\text{vec}(\textbf{V}_x);\text{vec}(\textbf{V}_y); \text{vec}(\bm{\theta}_{g_x}); \text{vec}(\bm{\theta}_{g_y})],
\end{aligned}
\end{multlined}
\end{equation}
where vec(.) denotes the operation of stacking all matrix elements into a column vector. The parameters of the univariate functions are denoted $\bm{\theta}_{g_x}$ and $\bm{\theta}_{g_y}$, respectively for the state and the output equation.

For completeness the cost function is repeated
\begin{equation}
\label{e:VLSbis}
\textbf{V}_{\text{LS}}(\bm{\theta}_{ss}) =  \frac{1}{N}\sum_{k=1}^N  \left\| \textbf{y}_{\text{meas}}(k) - \textbf{y}(\bm{\theta}_{ss},k)\right\|^2_2, \tag{\ref{e:VLS}}
\end{equation}
\revv{where $N$ is the number of samples, $\textbf{y}_{\text{meas}}$ is the true output and $\textbf{y}$ is the modelled output.}

As error metric of the model a relative root-mean-square value is used,
%
%
\begin{equation}
\label{e:e_rms}
e_{\text{rms}} = \sqrt{\frac{\frac{1}{N} \sum_{k=1}^N \norm{\textbf{y}_{\text{meas}}(k)-\textbf{y}(k)}_2^2}{\frac{1}{N} \sum_{k=1}^N \norm{\textbf{y}_{\text{meas}}(k)}_2^2}}.
\end{equation}

%

\section{Numerical and experimental case studies}
\label{s:case_studies}

In this section the model reduction approach is applied to a number of systems, both real-life and numerical.

\subsection{The forced Duffing oscillator}
\label{ss:silverbox}

This is an experimental case study on an electrical implementation of a mechanically resonating system involving a moving mass $m$, a viscous damping $c$ and a nonlinear spring $k(y(t))$. The analogue electrical circuitry generates data close to but not exactly equal to the idealised representation given by the nonlinear ordinary differential equation (ODE)
\begin{equation}
\label{e:diff_SB}
m \ddot{y}(t) + c \dot{y}(t) + k(y(t))y(t) = u(t),
\end{equation}
where the presumed displacement, $y(t)$, is considered the output and the presumed force, $u(t)$, is considered the input. Overdots denote the derivative with respect to time. The static position-dependent stiffness is given by
\begin{equation}
\label{e:SB_kNL}
k(y(t))=\alpha+\beta y^2(t).
\end{equation}

Rewriting Eq.~\eqref{e:diff_SB} by introducing $\textbf{x}(t) = \left[ y(t) \quad \dot{y}(t) \right]^{\text{T}}$ as state variables results in
\begin{subequations} \label{e:PNLSS_SB}
    \begin{empheq}[left={\empheqlbrace\,}]{align}
      & \dot{\textbf{x}}(t)=\left[ \begin{matrix} 0 \quad 1 \\ \rfrac{-c}{m} \quad \rfrac{-\alpha}{m} \end{matrix} \right] \textbf{x}(t) + \left[ \begin{matrix} 0\\\rfrac{1}{m} \end{matrix} \right] u(t) + \color{red} \left[ \begin{matrix} 0 \\ \rfrac{-\beta}{m} \end{matrix} \right] x^3_1(t) \label{e:PNLSS_SB_a}\\ 
      & y(t) = \left[1 \quad 0 \right] \textbf{x}(t). \label{e:PNLSS_SB_b}
    \end{empheq}
\end{subequations}
Notice from the nonlinear part (indicated in red) that this is inherently a 1-branch model containing a nonlinear feedback loop over one of the state variables. 

\subsubsection{Data set}

\textbf{The training data} consists of 10 successive realisations of a random odd multisine signal given by
\begin{equation}
\label{e:multisine}
u(t) = A \sum_{\begin{matrix} l=1\\l~odd \end{matrix}}^{L}\cos \left( 2\pi f_0 \ell t+ \phi_l \right),
\end{equation}
The period of the multisine is $\rfrac{1}{f_0}$ with $f_0= \rfrac{f_s}{8192}$ Hz and $f_s\approx 610$ Hz. The number of excited harmonics is $L=1342$ resulting in an $f_{\text{max}}\approx 200$ Hz. Each multisine realisation is given a unique set of phases $\phi_l$ that are independent and uniformly distributed in $[0,2\pi[$. The signal to noise ratio at the output is estimated at approximately 40 dB. 

As \textbf{validation data} two sets are available: an additional phase realisation of identical frequency range and a filtered Gaussian noise sequence of the same band width and with a linearly increasing amplitude to which we will refer as the arrow.

This data are part of three benchmark data sets for nonlinear system identification described in \cite{wigren2013}\revv{, and are used in a number of works among which \cite{noel2018,ljung2004,sragner2004}.}

\subsubsection{Results}

\begin{figure}
\center
\begin{subfigure}[b]{0.45\textwidth}
\begin{center}
\includegraphics[width=\textwidth]{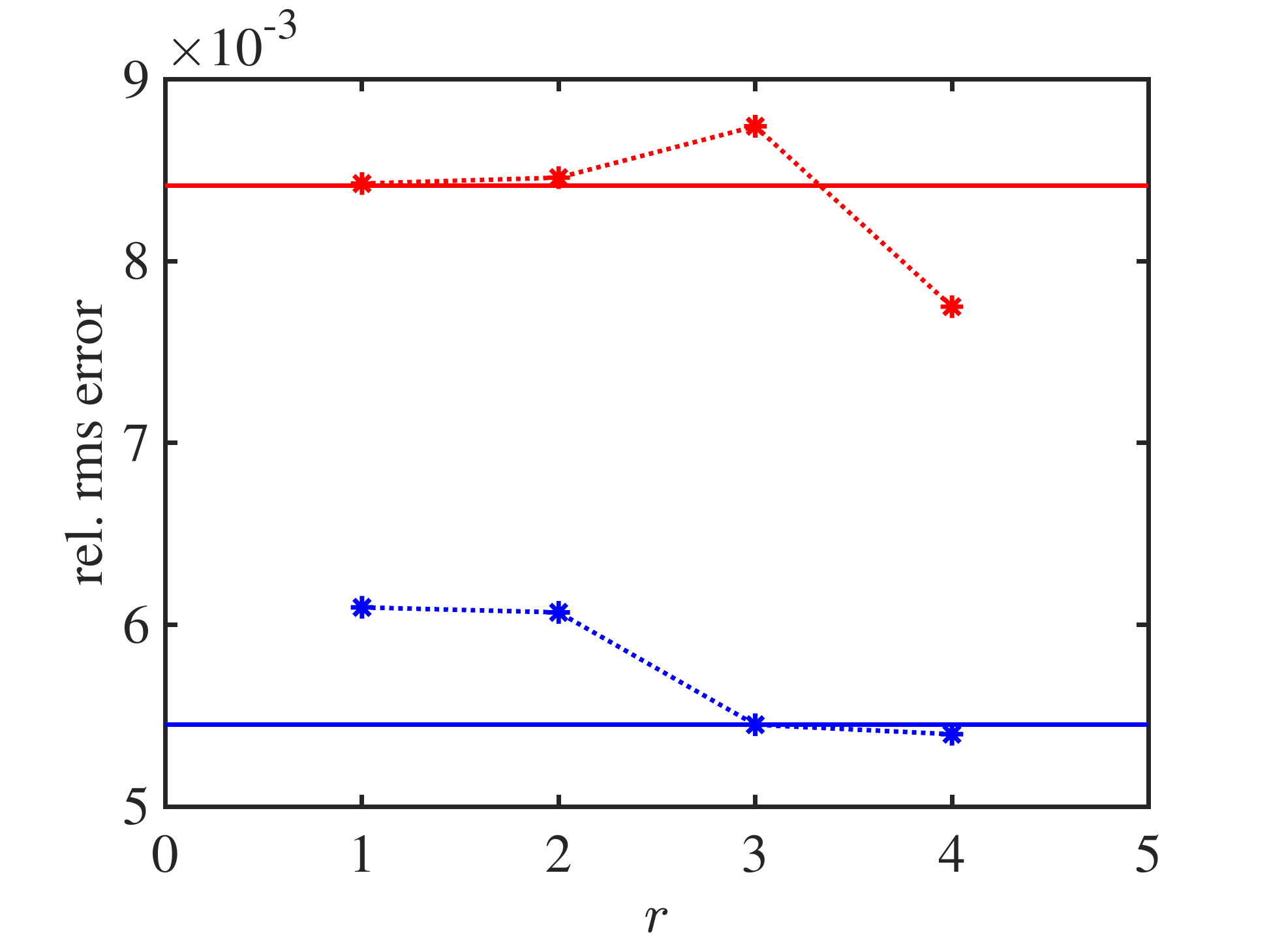}
\caption{}
\label{f:error_branches_SB}
\end{center}
\end{subfigure}
\begin{subfigure}[b]{0.45\textwidth}
\begin{center}
\includegraphics[width=\textwidth]{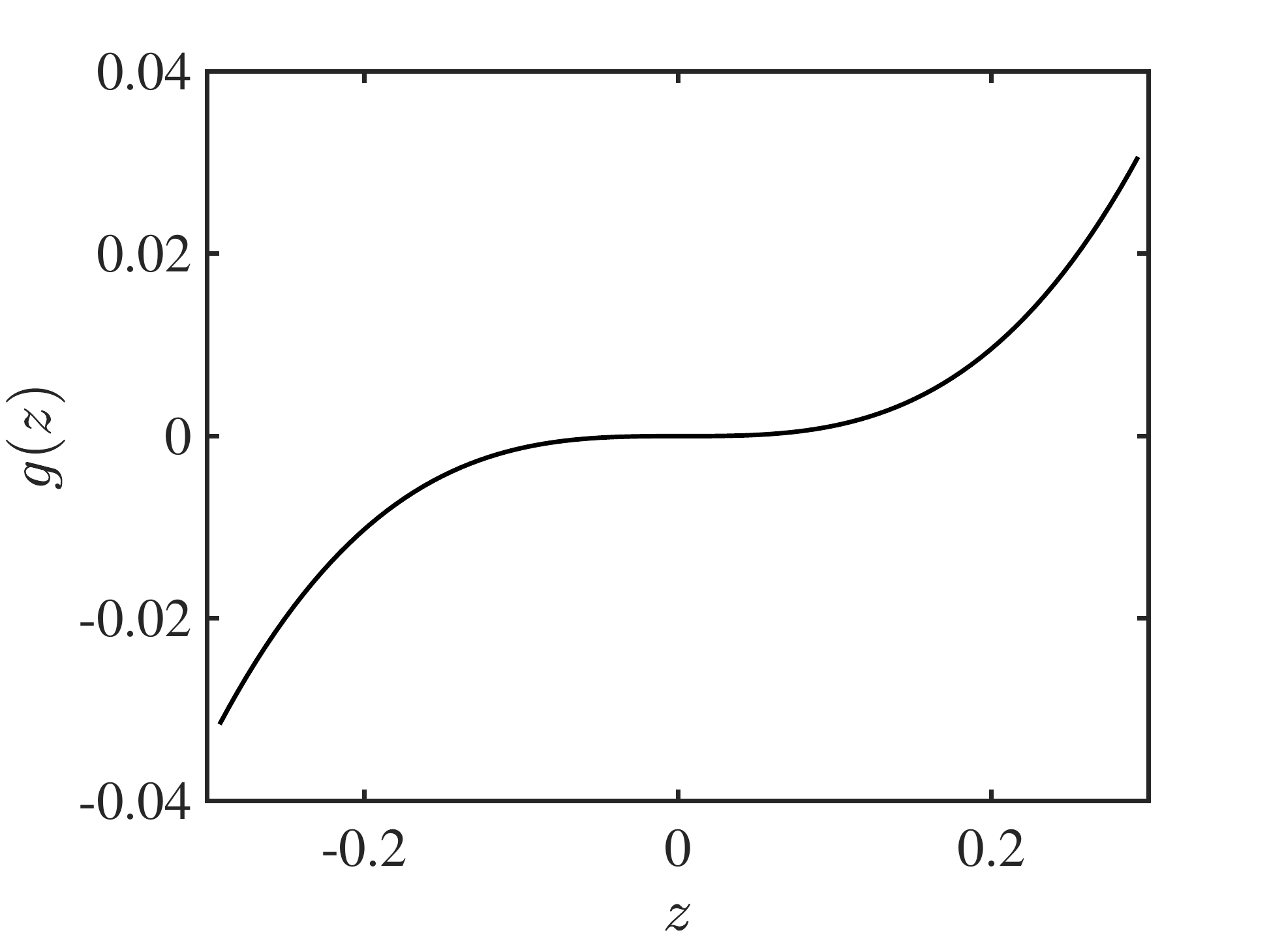}
\caption{}
\label{f:r1_SB}
\end{center}
\end{subfigure}
\caption{(a) Relative rms.\ error of the forced Duffing model as a function of the number of branches of the reduced model. Red corresponds to the results on the `arrow' validation set while blue is the additional multisine realisation. The markers show the results of the reduced models while the solid lines correspond to the original PNLSS model. (b) Visualisation of the single branch.}
\end{figure}

The PNLSS model was originally decoupled into $r=4$ branches following $r = \text{rank}~\mathcal{J}_x$. The performance of the reduced models is summarised in Fig.~\ref{f:error_branches_SB}. \revv{Reduced models with an equivalent performance with respect to the reference PNLSS model were obtained for certain values of $r$. The model with a single remaining branch performs equally well as the PNLSS model on \revv{the `arrow' validation set while performing slightly worse (although only 0.05\%) on the multisine validation set}.  \revv{Slight deviations with respect to the reference may be attributed to local minima encountered during the optimisation}. Given the non-convex nature of Eq~\eqref{e:VLS}, a loss in model performance (due to local minima) can occur even when the underlying system falls exactly within the reduced model structure, as is the case for the forced Duffing system. When decoupling the nonlinearity, unifying the branches and reducing their number to one, the number of degrees of freedom is almost halved, while the rms validation errors stay in the same ballpark (Table~\ref{t:results_SB}).}

\revv{The shape of the single remaining branch is shown in Fig.~\ref{f:r1_SB}. For this specific case, physical interpretation can be given to the nonlinearity. Section \ref{ss:class} provides a procedure from which we are able to infer that the underlying system behaves as a hardening spring.}

\begin{table}[h!]
\caption{Model reduction results of the forced Duffing system.}
\label{t:results_SB}
\begin{center}
\begin{tabular}{| c | c | c |}
\cline{2-3}
\multicolumn{1}{c|}{} & coupled PNLSS & $r=1$  \\
\hline
number of inputs to the nonlinearity & 2 & 2 \\
state nonlinearity $\textbf{f}_x$ &  degrees 2, 3 & degrees 2, 3\\
output nonlinearity $\textbf{f}_y$ & - & - \\
\# DOF & 19 & 10 \\
$e_{\text{rms}}$ validation realisation & 0.0085 & 0.0084\\
$e_{\text{rms}}$ validation arrow & 0.0054 & 0.0061\\
\hline
\end{tabular}
\end{center}
\end{table}

\begin{figure}
\begin{subfigure}[b]{\textwidth}
\begin{center}
\includegraphics[width=\textwidth]{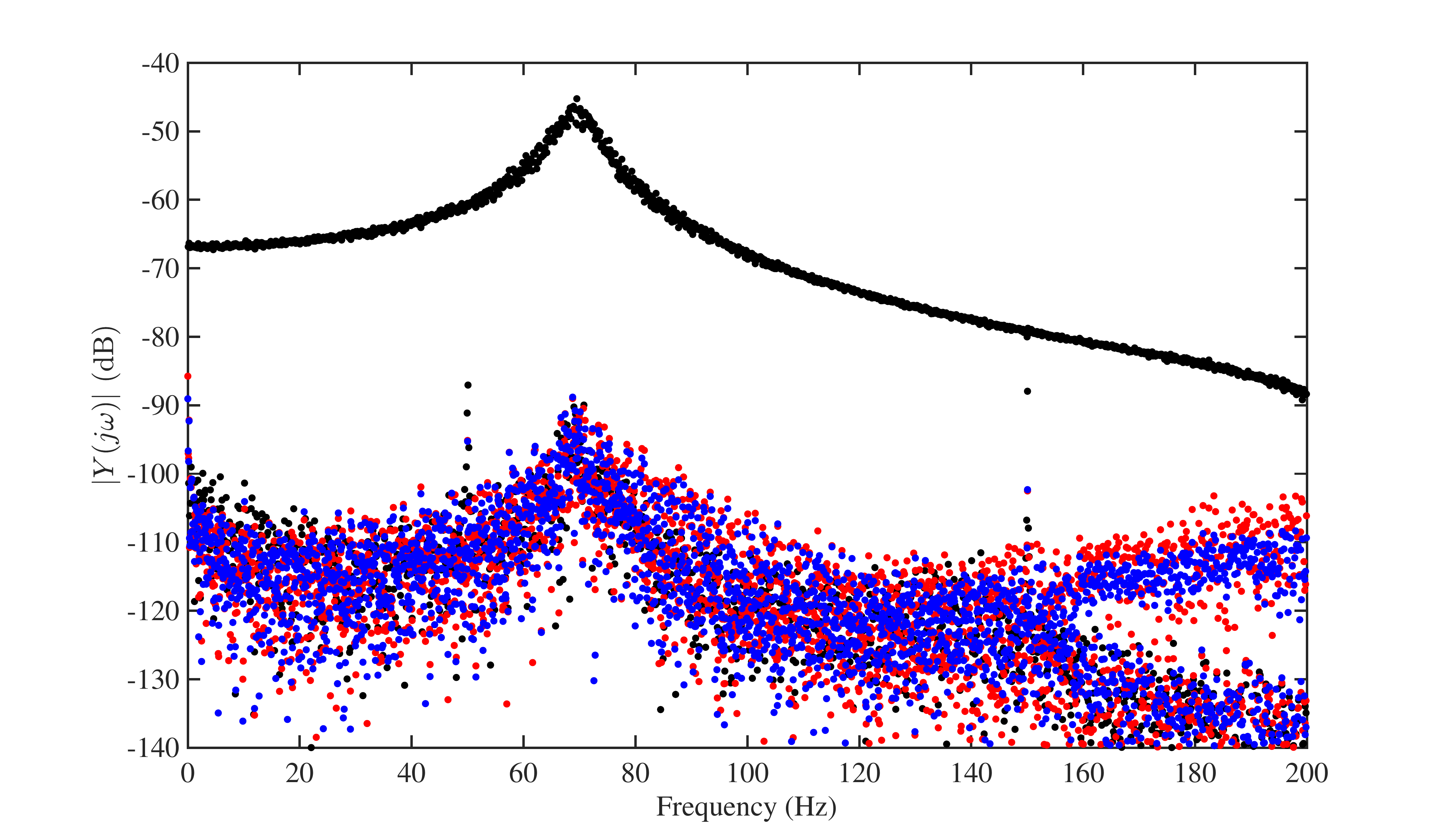}
\caption{}
\label{f:val_R_SB}
\end{center}
\end{subfigure}
\begin{subfigure}[b]{\textwidth}
\begin{center}
\includegraphics[width=\textwidth]{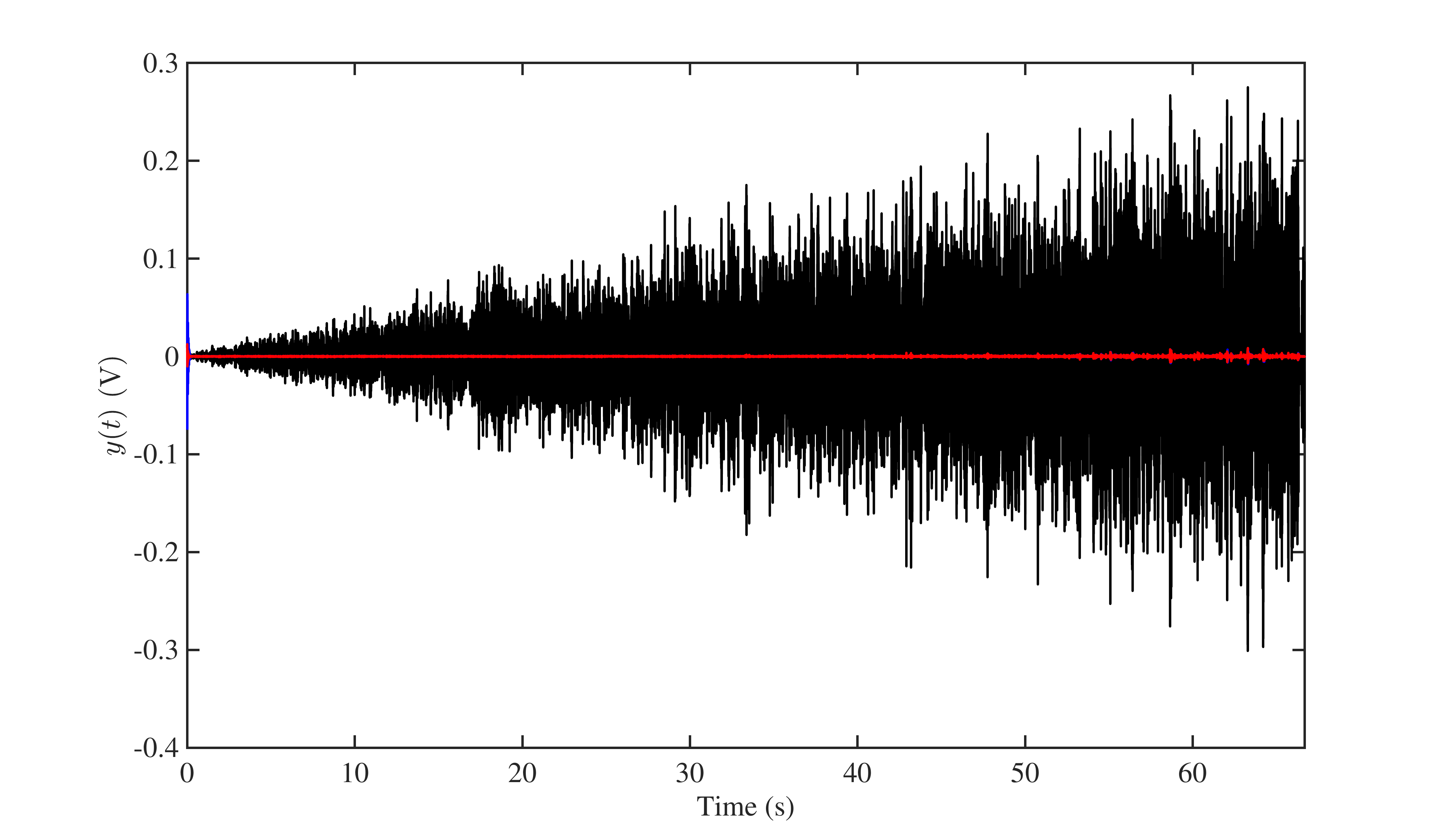}
\caption{}
\label{f:val_arrow_SB}
\end{center}
\end{subfigure}
\caption{(a) Spectrum of the validation results of the 1-branch forced Duffing model. (b) Validation error on the Gaussian noise arrow data set. In both panels black shows the output data, blue is the error of the original coupled PNLSS model and red is the error of the 1-branch decoupled model. }
\end{figure}

\revv{Fig.\ \ref{f:val_R_SB} shows the validation results of the 1-branch model on the additional multisine realisation. The results on the arrow \rev{noise sequence} are shown in Fig.~\ref{f:val_arrow_SB}. The figures reflect what was concluded from Table \ref{t:results_SB}, i.e.\ a similar performance is achieved using the one-branch model. Both models are able of accurately capturing the response in the frequency band of interest. In Fig.\ref{f:val_arrow_SB}, it is interesting that the error stays relatively small at the end of the experiment where there is extrapolation of the model. The test signal has amplitudes exceeding those present in the multisine training signal. Typically, black-box models have difficulties with these extrapolations.} 

\revv{To stress the level of reduction that was achieved, the fully coupled PNLSS model and its decoupled (reduced) counterpart are written in full below:
\begin{subequations} \label{e:SB1}
    \begin{empheq}[left={\empheqlbrace\,}]{align}
      & \textbf{x}(k+1)=\textbf{A}\textbf{x}(k)+\textbf{b}u(k)+ \color{red} \left[ \begin{matrix} e_{11} \quad e_{12} \quad e_{13} \quad e_{14} \quad e_{15} \quad e_{16} \quad e_{17}  \\ e_{21} \quad e_{22} \quad e_{23} \quad e_{24} \quad e_{25} \quad e_{26}\quad e_{27} \end{matrix} \right]
       \left[ \begin{matrix} x_1^2(k) \\ x_1(k)x_2(k) \\ x_2^2(k) \\ x_1^3(k) \\ x_1^2(k)x_2(k) \\ x_1(k)x_2^2(k) \\ x_2^3(k) \end{matrix} \right] \\ \color{black}
      & y(k)=\textbf{c}^{\text{T}}\textbf{x}(k)+du(k),
\end{empheq}
\end{subequations}
with 19 DOF of which 10 correspond to the nonlinear part. And the reduced decoupled model for $r=1$:
\begin{subequations} \label{e:SB2}
    \begin{empheq}[left={\empheqlbrace\,}]{align}
      & \textbf{x}(k+1)=\textbf{A}\textbf{x}(k)+\textbf{b}u(k)+ \color{red} \left[ \begin{matrix} w_1  \\ w_2  \end{matrix} \right] \left[ \theta_{1}z^3(k)+ \theta_{2}z^2(k)\right]  \color{black} \label{e:SB2a}\\ 
      & y(k)=\textbf{c}^{\text{T}}\textbf{x}(k)+du(k),\\
      & z(k) = [v_1 \quad v_2] \left[ \begin{matrix} x_1(k) \\ x_2(k) \end{matrix} \right],
\end{empheq}
\end{subequations}
having only 10 DOF of which only 3 are used in the description of the nonlinearity.}

\revv{\subsubsection{Intermediate conclusions}
The forced duffing benchmark illustrates how overly complex models are found using the classical PNLSS structure. In this case the underlying system is inherently described by a one-branch nonlinear function. Decoupling the polynomial and consecutive application of the branch reduction step enables to retrieve this single-branch form.}

\subsection{The Van der Pol oscillator}
\label{ss:VdP}

This is a numerical case study of the Van der Pol oscillator, described by a second order nonlinear ODE with a nonlinear damping term. In reduced form it reads
\begin{equation}
\label{e:diff_VdP}
\ddot{y}(t) + \varepsilon\left(y^2(t)-1 \right) \dot{y}(t) + \omega_0^2y(t) = u(t),
\end{equation}

where $u(t)$ is a forcing term, $y(t)$ is considered the output \cite{vanderpol1926} and $\omega_0$ sets the angular natural frequency. A weight on the nonlinear term is provided through the Van der Pol parameter $\varepsilon$. Having a nonlinear damping term, able to introduce both negative and positive damping depending on the output level, results in regimes of autonomous oscillation and limit cycle amplitudes. These properties are of particular interest when modelling unsteady fluid dynamics \cite{dowell1981,hartlen1970,parkinson1974}.

Rewriting Eq.~\eqref{e:diff_VdP} by introducing $\textbf{x}(t) = \left[ y(t) \quad \dot{y}(t) \right]^{\text{T}}$ as state variables results in
\begin{subequations} \label{e:PNLSS_VdP}
    \begin{empheq}[left={\empheqlbrace\,}]{align}
      & \dot{\textbf{x}}(t)=\left[ \begin{matrix} 0 \quad 1 \\ -\omega_0^2 \quad \varepsilon \end{matrix} \right] \textbf{x}(t) + \left[ \begin{matrix} 0\\1 \end{matrix} \right] u(t) + \color{red} \left[ \begin{matrix} 0 \\-\varepsilon \end{matrix} \right] x^2_1(t) x_2(t) \label{e:PNLSS_VdP_a}\\ 
      & y(t) = \left[1 \quad 0 \right] \textbf{x}(t), \label{e:PNLSS_VdP_b}
    \end{empheq}
\end{subequations}
where the nonlinear part is highlighted in red. It is clear from the cross-term monomial that no exact decoupling into a 1-branch function exists. Eq.~\eqref{e:exact_dec} shows that at least 3 branches are needed to decouple the given term.

\subsubsection{Data set}

The data are generated using a first order forward Euler discretisation of Eq.~\eqref{e:PNLSS_VdP},
\begin{subequations} \label{e:VdP_disc}
    \begin{empheq}[left={\empheqlbrace\,}]{align}
      & \textbf{x}(k+1)=\left[ \begin{matrix} 1 \quad T_s \\ -\omega_0^2T_s \quad \varepsilon T_s+1 \end{matrix} \right] \textbf{x}(k) + \left[ \begin{matrix} 0\\T_s \end{matrix} \right] u(k) + \color{red} \left[ \begin{matrix} 0 \\-\varepsilon T_s \end{matrix} \right] x^2_1(k) x_2(k) \label{e:VdP_disc_a}\\ 
      & y(k) = \left[1 \quad 0 \right] \textbf{x}(k), \label{e:VdP_disc_b}
    \end{empheq}
\end{subequations}
with $k = \rfrac{t}{T_s}$, a sample period $T_s = 0.01s$, and with parameters: $\varepsilon = 0.03$, $\omega_0 = 2\pi$. This discrete-time PNLSS model is considered as the true underlying model.

\textbf{The training data} consists of 4 realisations of a random-phase multisine signal given by
\begin{equation}	
\label{e:multisine}
u(t) = A \sum_{l=1}^L \cos \left( 2\pi f_0 \ell t+ \phi_l \right),
\end{equation}
The period of the multisine is $\rfrac{1}{f_0}$ with $f_0= 0.01$ Hz. The number of excited harmonics is $L=400$ resulting in an $f_{\text{max}}=4$ Hz. The phases $\phi_l$ are independent and uniformly distributed in $[0,2\pi[$. The input was scaled such that $\text{rms}(\textbf{u})=50$. No noise was added to the output.

As \textbf{validation data} an additional phase realisation of identical frequency range and amplitude is used.

\subsubsection{Results}

\begin{figure}
\center
\begin{subfigure}[b]{0.45\textwidth}
\begin{center}
\includegraphics[width=\textwidth]{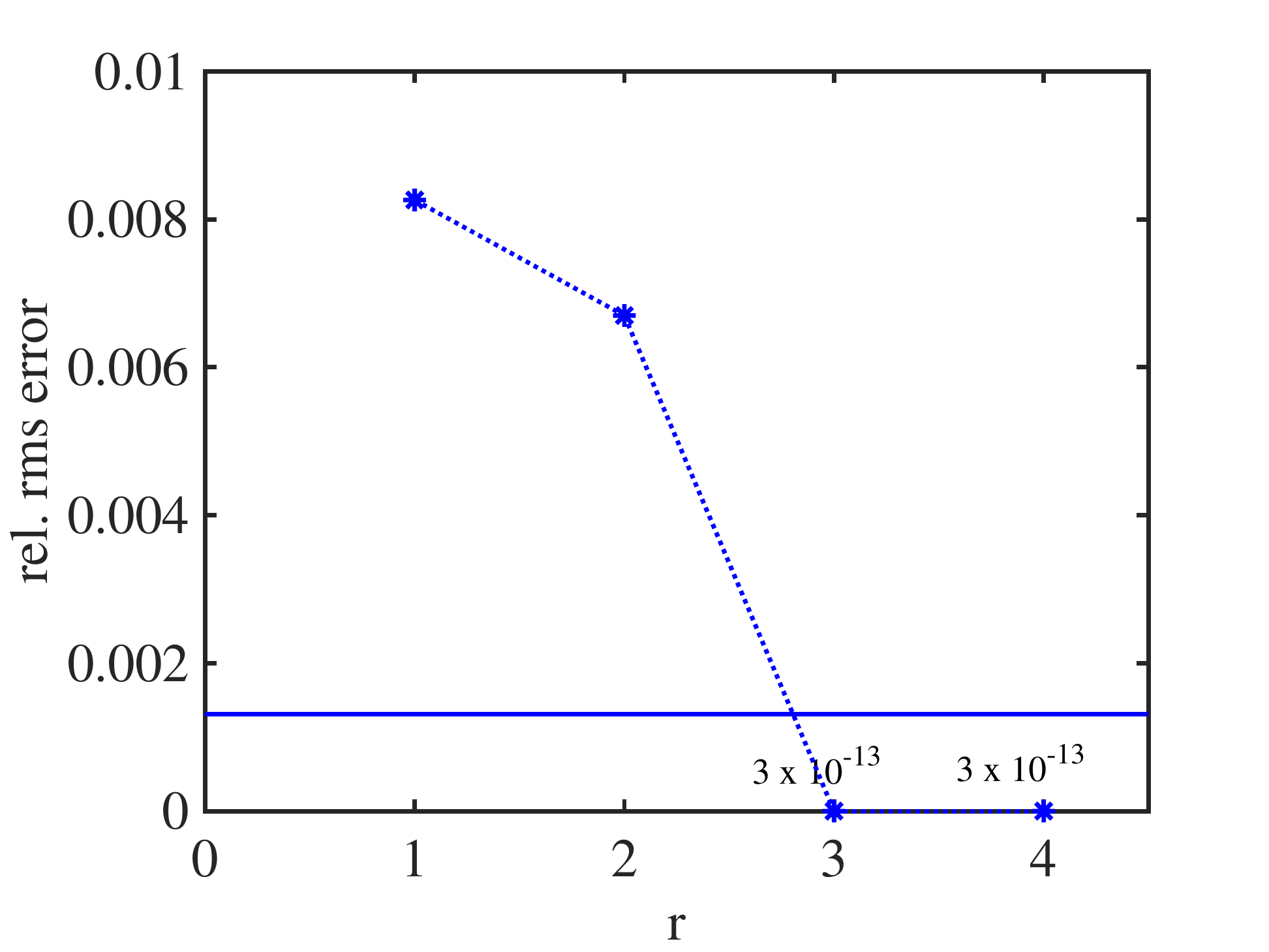}
\caption{}
\label{f:error_branches_VdP}
\end{center}
\end{subfigure}
\begin{subfigure}[b]{0.45\textwidth}
\begin{center}
\includegraphics[width=\textwidth]{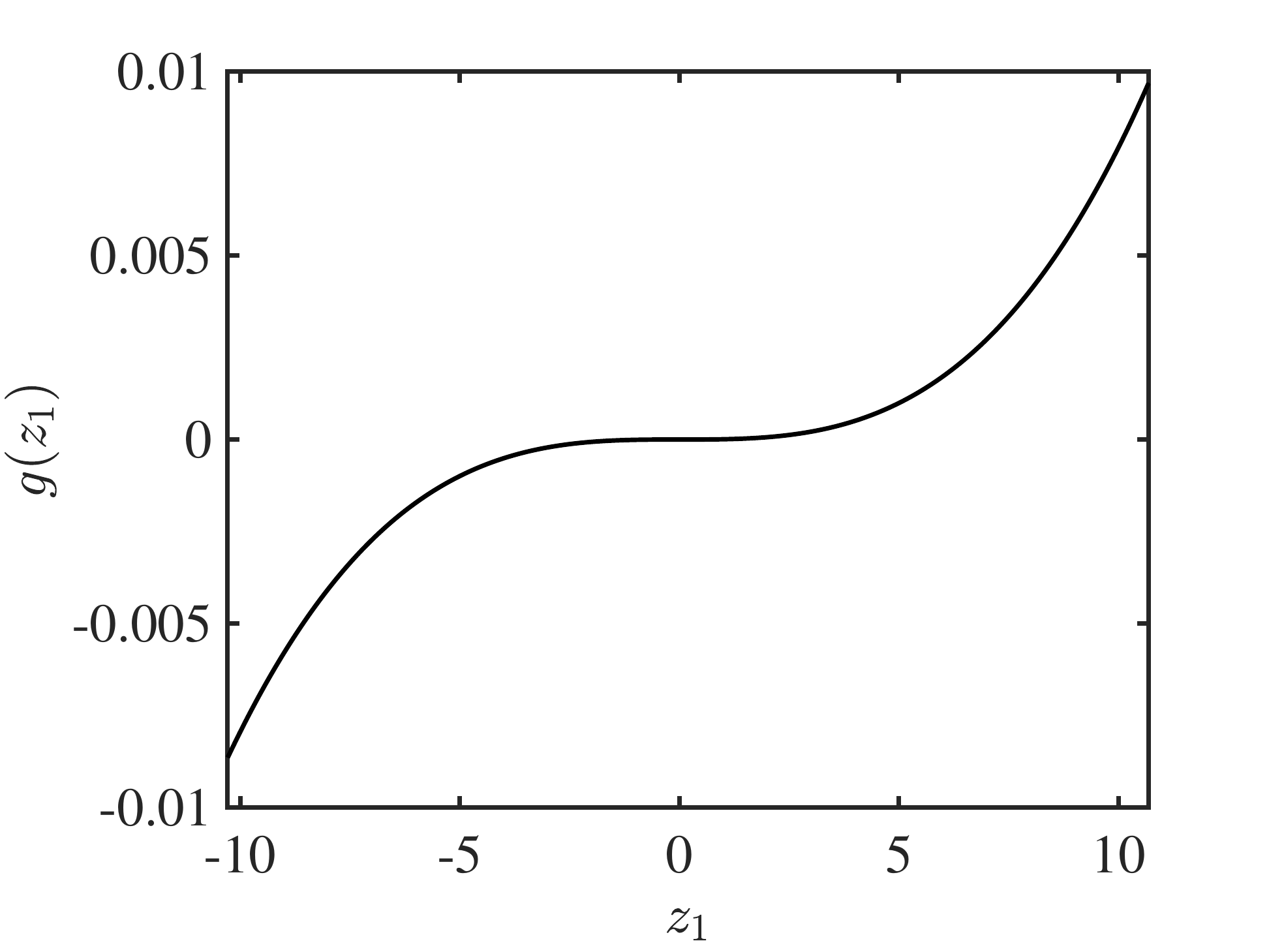}
\caption{}
\label{f:r1_VdP}
\end{center}
\end{subfigure}
\caption{(a) Relative rms.\ error on the validation data of the Van der Pol system as a function of the number of branches in the reduced model. Dots are the results of the reduced model while the solid line corresponds to the original PNLSS model. (b) Visualisation of the single branch.}
\end{figure}

\begin{figure}
\begin{center}
\includegraphics[width=\textwidth]{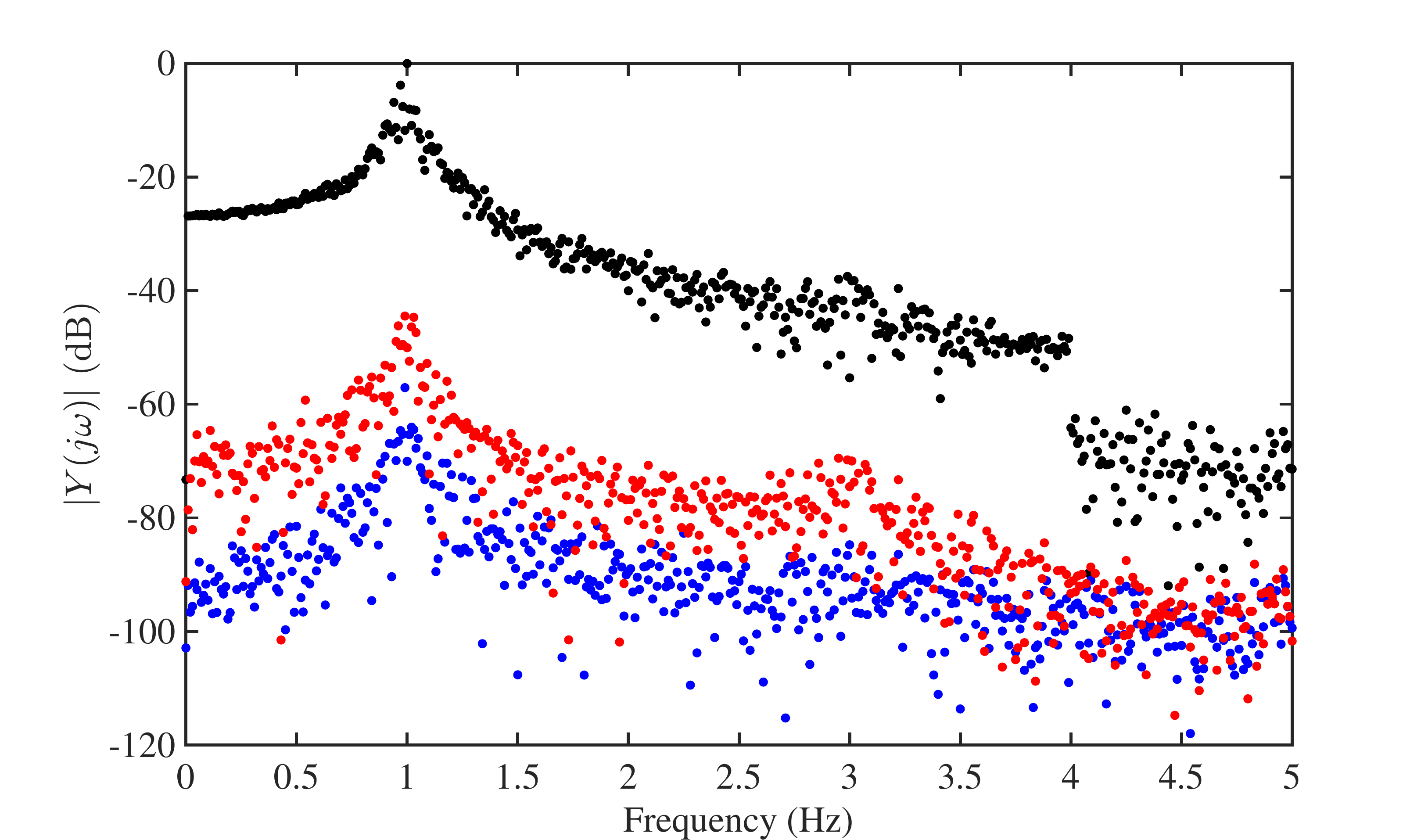}
\caption{Spectrum of the validation results of the 1-branch Van der Pol model.  Black shows the output data, blue is the error of the original coupled PNLSS model and red is the error of the 1-branch decoupled model. Recall that the Van der Pol system does not admit a 1-branch decoupled form. Exact solutions can only be found for values of $r\ge3$.}
\label{f:val_R_results_VdP}
\end{center}
\end{figure}

\rev{The static nonlinearity $\textbf{f}_x$ in the PNLSS model was originally decoupled into $r=4$ branches. The validation results of the reduced models are summarised in Fig.~\ref{f:error_branches_VdP} and additionally listed in Table~\ref{t:results_VdP}. Notice that the PNLSS model still shows an error even though the Van der Pol system of Eq.~\eqref{e:PNLSS_VdP} falls exactly within the PNLSS model class and no noise was added to the data. This again illustrates the impact of local minima, attained during the optimisation. 

Given the nature of the nonlinearity ($x_1^2x_2$), we know that an exact decompositions exist for values of $r\ge3$ (see Eq.~\eqref{e:exact_dec}). Such decoupled form is retrieved by the method resulting in validation errors close to machine precision (Fig.~\ref{f:error_branches_VdP}). Notice that for the decoupled form an accurate local minimum can be found. Reducing the number of branches below $r=3$ inevitably introduces errors. \revv{The function which is retrieved for a single-branch model is depicted in Fig.~\ref{f:r1_VdP}. Physical interpretation is sought in Section \ref{ss:class}. Both nonlinear stiffness and nonlinear damping are associated to the behaviour of the system in this case. Fig.~\ref{f:val_R_results_VdP} depicts the error of the approximate Van der Pol model. It illustrates that a one-branch model introduces errors which are at the 1\% level (20 dB lower than the output level) in the frequency band of interest.}  }

\begin{table}[h!]
\caption{Model reduction results of the Van der Pol system.}
\label{t:results_VdP}
\begin{center}
\begin{tabular}{| c | c | c |}
\cline{2-3}
\multicolumn{1}{c|}{} & coupled PNLSS & $r=1$  \\
\hline
number of inputs to the nonlinearity & 2 & 2 \\
state nonlinearity $\textbf{f}_x$ &  degrees 3 & degree 3\\
output nonlinearity $\textbf{f}_y$ & - & - \\
\# DOF & 13 & 8 \\
$e_{\text{rms}}$ validation realisation & 0.0013 & 0.0082 \\
\hline
\end{tabular}
\end{center}
\end{table}

\revv{\subsubsection{Intermediate conclusions}

From the Van der Pol system we learn that even in the exact case, when the underlying system falls exactly within the model class (by construction), the classical PNLSS structure may suffer from the non-convex nature of the cost function, resulting in model errors. Although the decoupled form also faces a non-convex cost, the optimisation landscape is altered by the decoupling, potentially resulting in improved models.}

\subsection{The Bouc-Wen hysteresis system}
\label{ss:Bouc_Wen}

This is a numerical case study of a Bouc-Wen model realisation. The Bouc-Wen model has been intensively used to represent hysteretic effects in mechanical engineering \cite{morrison2001,bertotti1998,mueller1985}. The defining hysteresis loop follows from a nonlinear memory-dependent restoring force ($f_{H}$). The dynamics of a single-degree-of-freedom Bouc-Wen oscillator are governed by the second order nonlinear ODE,
\begin{equation}
\label{e:diff_BW}
m\ddot{y}(t) + c\dot{y}(t) + ky(t) + f_{H}(y(t),\dot{y}(t)) = u(t),
\end{equation}
where $k$ and $c$ are the linear stiffness and viscous damping coefficients, respectively. The hysteretic force $f_{H}$ obeys the first order ODE
\begin{equation}
\label{e:diff_BW_2}
\dot{f}_{H}(t) = \alpha \dot{y}(t) - \left( \gamma |\dot{y}(t)||f_{H}(t)|^{\nu-1}f_H(t)+ \delta \dot{y}(t) |f_H(t)|^{\nu} \right),
\end{equation} 
with the Bouc-Wen parameters $\alpha$, $\beta$, $\gamma$, $\delta$ and $\nu$.

Rewriting Eq.~\eqref{e:diff_BW} by introducing $\textbf{x}(t) = \left[ y(t) \quad \dot{y}(t) \quad f_H(t) \right]^{\text{T}}$ as state variables and setting $\nu = 1$ results in
\begin{subequations} \label{e:PNLSS_BW}
    \begin{empheq}[left={\empheqlbrace\,}]{align}
   & \dot{\textbf{x}}(t)=\left[ \begin{matrix} 0 \quad 1 \quad 0 \\ \rfrac{-k}{m} \quad \rfrac{-c}{m} \quad \rfrac{-1}{m} \\ 0 \quad \alpha \quad 0 \end{matrix} \right] \textbf{x}(t) + \left[ \begin{matrix} 0\\\rfrac{1}{m} \\ 0 \end{matrix} \right] u(t) + \color{red} \left[ \begin{matrix} 0 \quad 0 \\ 0 \quad 0 \\ -\gamma \quad -\delta \end{matrix} \right] \left[ \begin{matrix} |x_2(t)| x_3(t) \\ x_2(t) |x_3(t)| \end{matrix} \right] \label{e:PNLSS_BW_a}\\ 
      & y(t) = \left[1 \quad 0 \right] \textbf{x}(t), \label{e:PNLSS_BW_b}
    \end{empheq}
\end{subequations}
where the nonlinear part is highlighted in red. Also here no exact decoupling into a 1-branch function exists. \revv{In fact, no exact PNLSS model exists, provided a finite set of monomial basis functions is used.}
\subsubsection{Data set}

The data are part of a benchmark for nonlinear system identification described in \cite{schoukens2017}. The parameter values are listed in Table \ref{t:BW}. 
\begin{table}[h!]
\caption{Parameter values of the Bouc-Wen model.}
\label{t:BW}
\begin{center}
\begin{tabular}{c  c  c  c c c  c c c }
\hline
Parameter & m & c & k & $\alpha$ & $\beta$ & $\gamma$ & $\delta$ & $\nu$ \\
Value (in SI unit) & 2 & 10 & 5 $10^4$ & 5 $10^4$ & 1 $10^4$ & 0.8 & -1.1 & 1 \\
\hline
\end{tabular}
\end{center}
\end{table}

\textbf{The training data} consist of 4 realisation of a multisine (Eq.~\eqref{e:multisine}) exciting all frequencies in the band from 5 - 150 Hz. The input level corresponds to $\text{rms}(\textbf{u})= 50$ N. The signal-to-noise ratio on the output is approximately 40 dB.

As \textbf{validation data} two data sets are used: an additional phase realisation of identical frequency range and amplitude and a sine sweep, sweeping from 5 Hz to 20 Hz at a sweep rate of 10 Hz/min and an amplitude level $\text{rms}(\textbf{u}) = 40$ N.

\subsubsection{Results}

\begin{figure}
\center
\begin{subfigure}[b]{0.3\textwidth}
\begin{center}
\includegraphics[width=\textwidth]{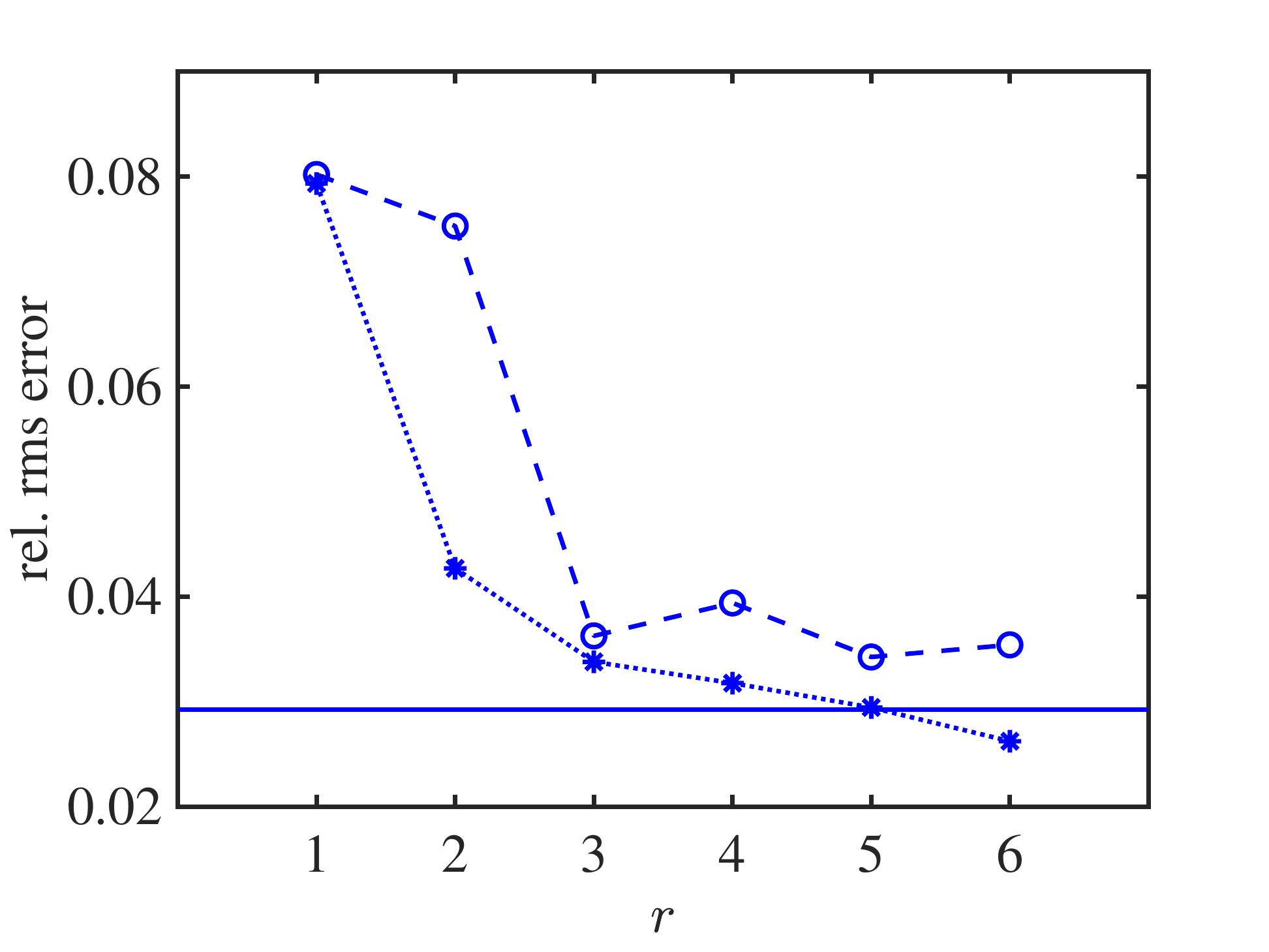}
\caption{}
\label{f:error_branches_uni_BW_a}
\end{center}
\end{subfigure}
\begin{subfigure}[b]{0.3\textwidth}
\begin{center}
\includegraphics[width=\textwidth]{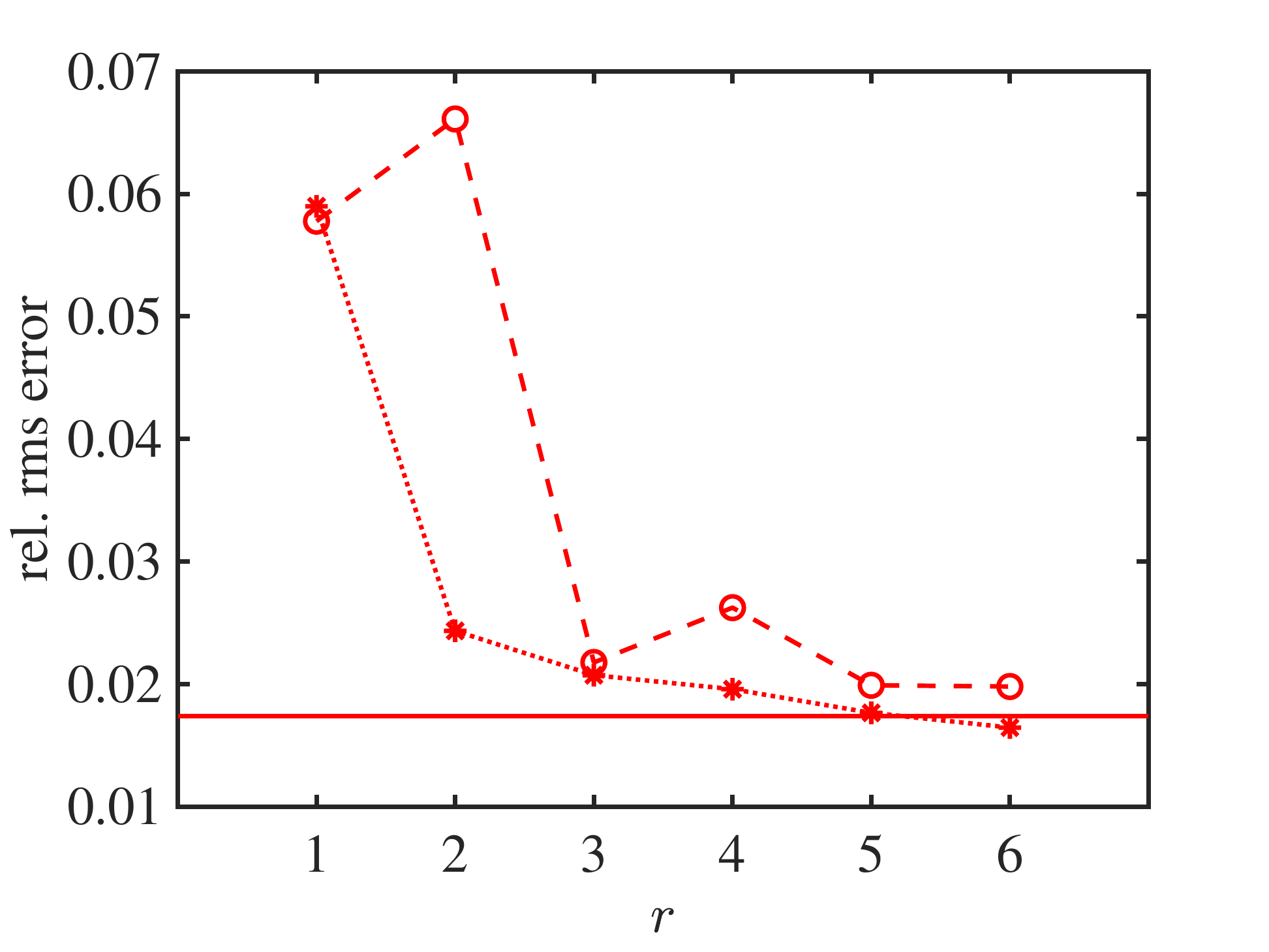}
\caption{}
\label{f:error_branches_uni_BW_b}
\end{center}
\end{subfigure}
\begin{subfigure}[b]{0.3\textwidth}
\begin{center}
\includegraphics[scale=0.222]{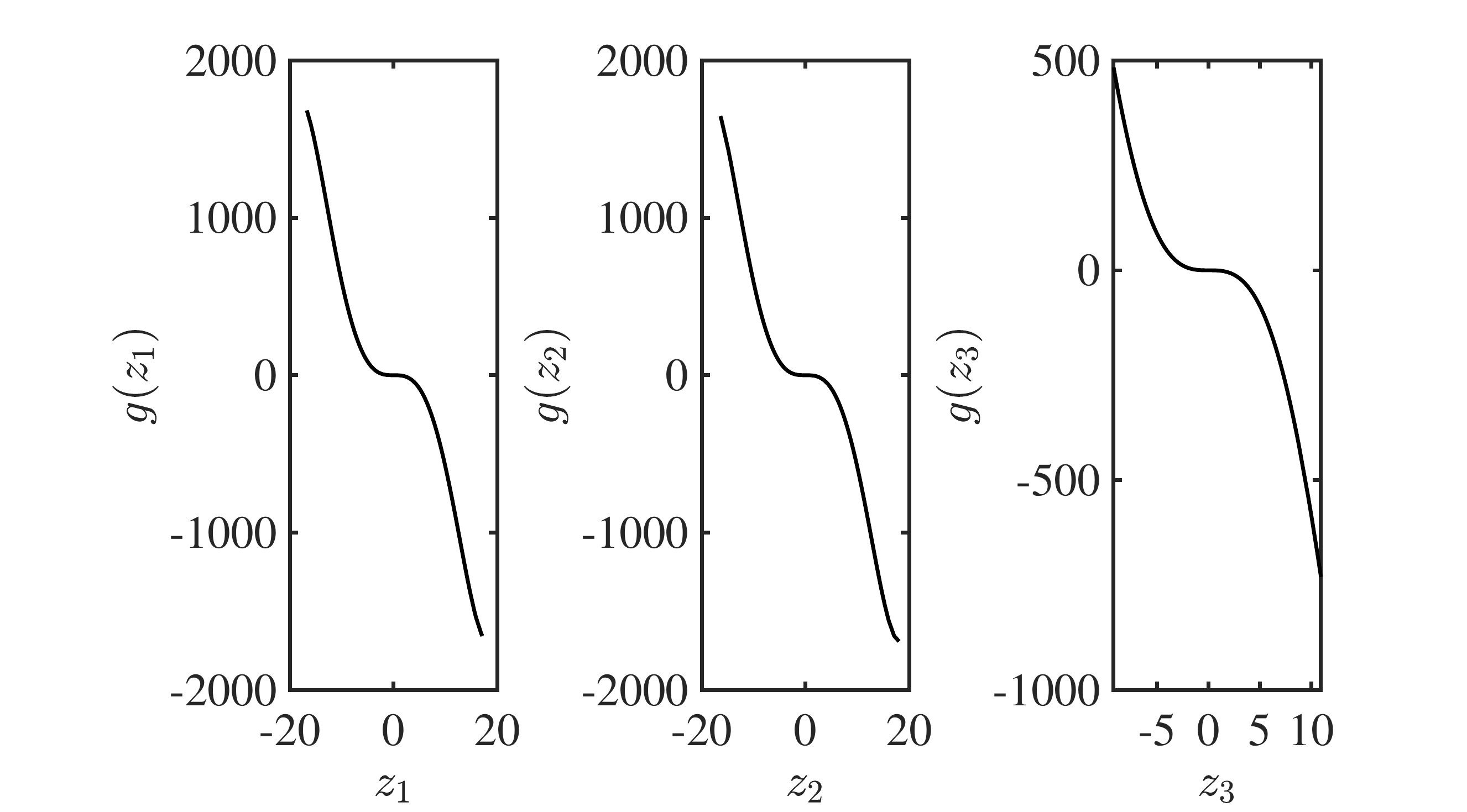}
\caption{}
\label{f:r1_BW}
\end{center}
\end{subfigure}
\caption{Validation results of the decoupled Bouc-Wen model as a function of number of branches. (a) Relative rms error on the multisine validation realisation. (b) Relative rms error on the swept sine validation data. Solid line correspond to the original coupled PNLSS model, `$\star$'-markers indicate the results of the reduced decoupled models, `o'-markers indicate the results of the decoupled and unified reduced models. (c) Evaluation of the three identical branches of the $r=3$ unified model.}
\end{figure}

\rev{The static function $\textbf{f}_x$ of the PNLSS model was originally decoupled into $r=6$ branches following $r = \text{rank}~\mathcal{J}_x$. For the present system it was found worthwhile to study both the reduced models and the reduced models with unified branches. The summary of the validation results for both validation data sets are presented in Fig.~\ref{f:error_branches_uni_BW_a} and Fig.~\ref{f:error_branches_uni_BW_b}. \revv{The results show} that a reduced model with three branches achieves a similar performance as the reference value of the coupled PNLSS model. Notice, moreover, that a model with $r=3$ and with unified branches is equally performant as the non-unified model whilst resulting in an additional decrease of the DOF. \revv{A visualisation of the unified 3-branch model nonlinearity is provided in Fig.~\ref{f:r1_BW}. No physical interpretation can be given to the nonlinearity in case of a multiple branch model (see Section \ref{ss:class}). One can however observe that the system behaves odd-nonlinear, i.e.\ the unified branches have an odd nonlinear function ($g(-z_i) = -g(z_i)$). This is in agreement with the nonlinear functions in Eq.~\eqref{e:PNLSS_BW_a}.}

The results, listed in Table \ref{t:results_BW}, show that higher (with respect to the coupled PNLSS model) polynomial degrees can be used in the decoupled structure \revv{since in this case the number of parameters grows linearly with the degree while coupled polynomials suffer from a combinatorial growth of the number of parameters.} When the reduction is continued down to $r=1$ the descriptive power of the model decreases. It is up to the user to balance model accuracy to complexity for the intended application. }

\begin{figure}
\begin{subfigure}[b]{\textwidth}
\begin{center}
\includegraphics[width=\textwidth]{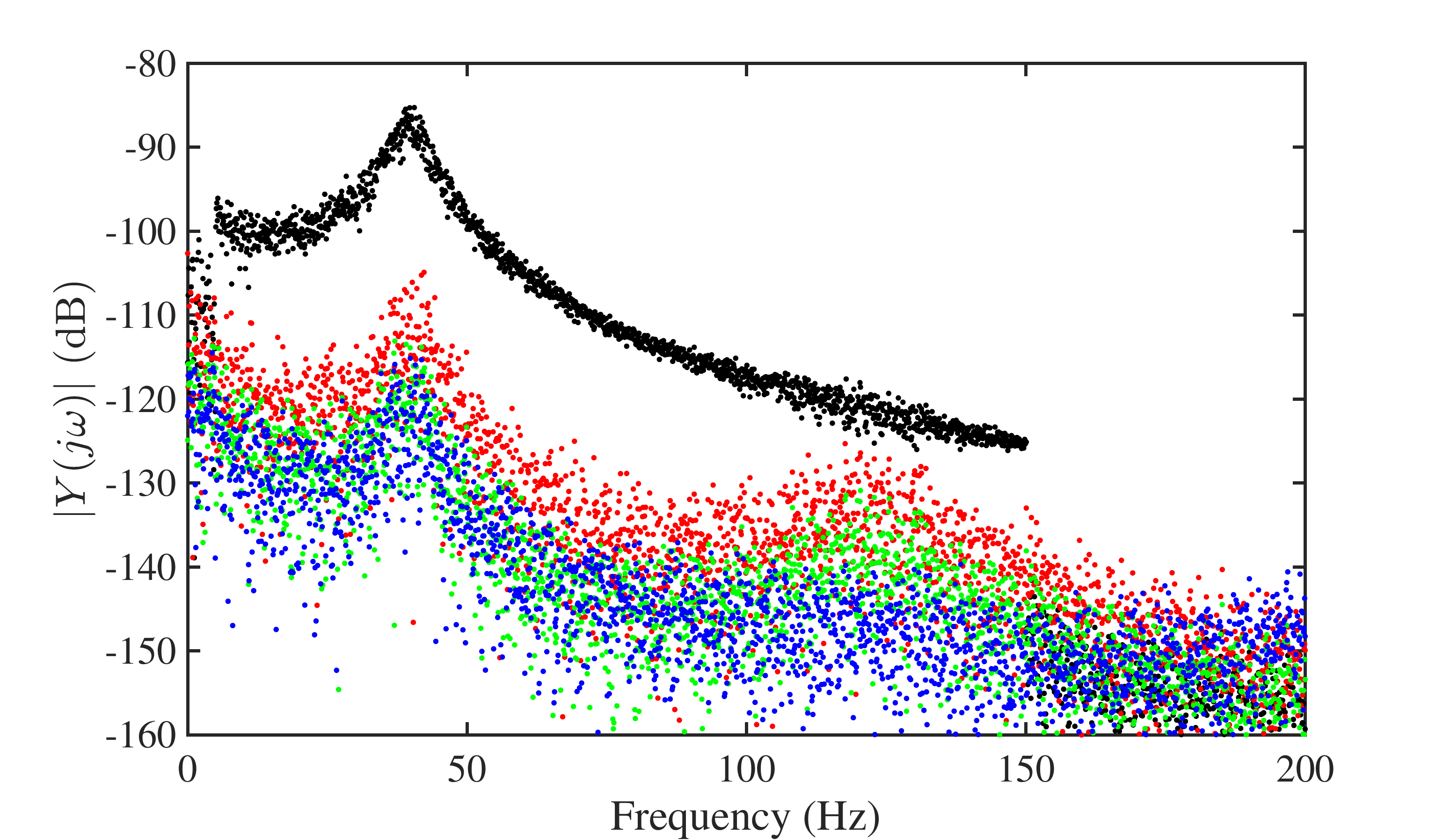}
\caption{}
\label{f:val_R_BW}
\end{center}
\end{subfigure}
\begin{subfigure}[b]{\textwidth}
\begin{center}
\includegraphics[width=\textwidth]{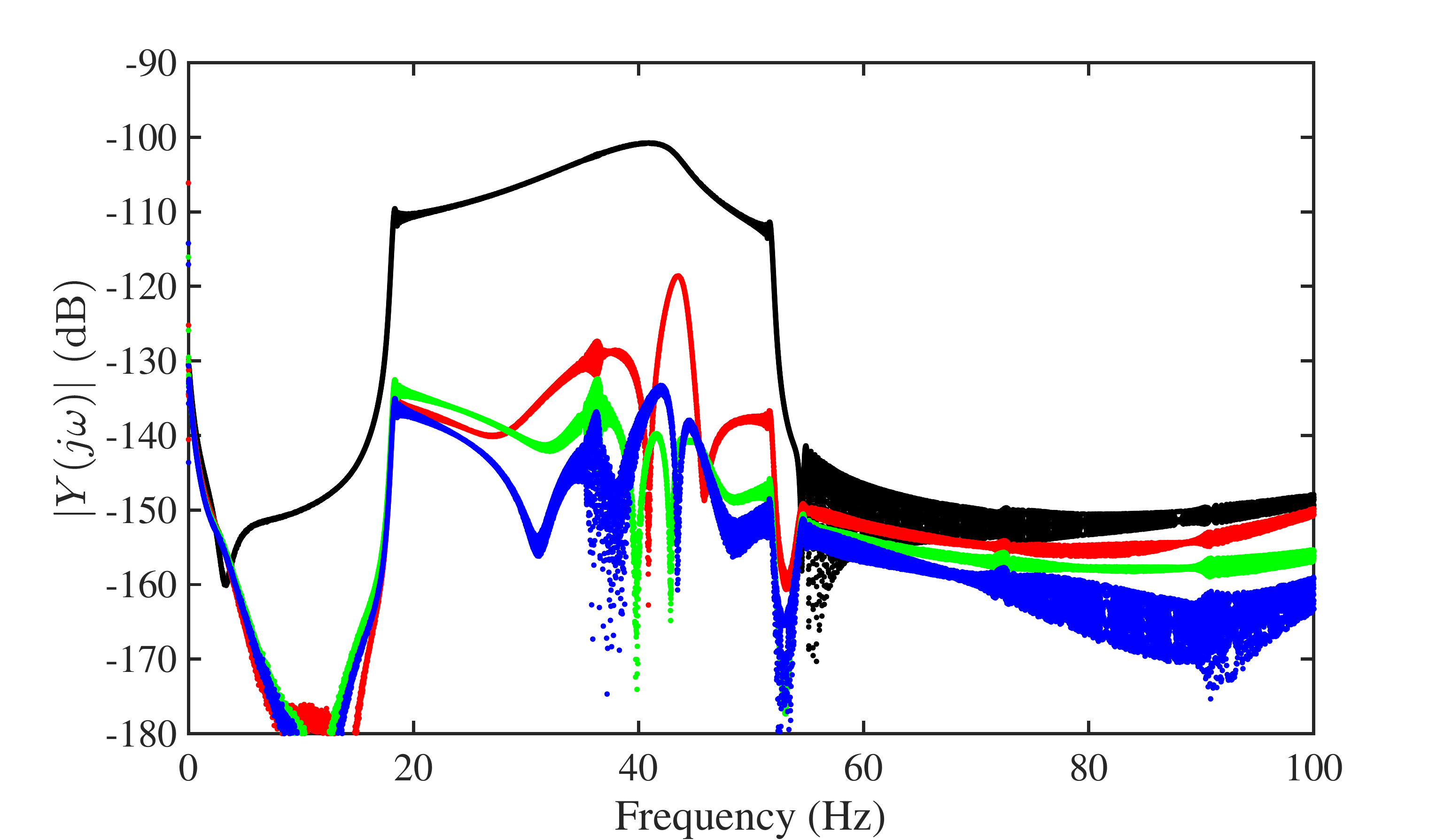}
\caption{}
\label{f:val_S_BW}
\end{center}
\end{subfigure}
\caption{Spectrum of the validation results of the reduced Bouc-Wen model. (a) Validation error on the multisine realisation. (b) Validation error on the sine sweep data. In both panels black shows the output data, blue is the error of the original coupled PNLSS model, red is the error of the 1-branch decoupled model and green is the error of the 3-branch unified decoupled model.}
\end{figure}

Frequency-domain plots of the validation realisation and the sine sweep validation are shown in Fig.~\ref{f:val_R_BW} and \ref{f:val_S_BW}, respectively. \revv{In case of the multisine, the figure shows that the reduced models introduce errors around the third harmonic of the resonance, hinting on the limitations of the nonlinear descriptive power of such reduced forms.}

\begin{table}[h!]
\begin{footnotesize}
\caption{Model reduction results of the Bouc-Wen system.}
\label{t:results_BW}
\begin{center}
\begin{tabular}{| c | c | c | c |}
\cline{2-4}
\multicolumn{1}{c|}{} & coupled PNLSS & $r=3$ unified & $r=1$  \\
\hline
number of inputs to the nonlinearity & 4 & 4 & 4 \\
state nonlinearity $\textbf{f}_x$ &  degrees 2,3 & degrees 2, 3, 4, 5 & degrees 2, 3, 4, 5 \\
output nonlinearity $\textbf{f}_y$ & - & - & - \\
\# DOF & 97 & 30 & 16 \\
$e_{\text{rms}}$ validation realisation & 0.0292 & 0.036  & 0.079 \\
$e_{\text{rms}}$ validation sine sweep & 0.0174 &  0.022 &  0.059 \\
\hline
\end{tabular}
\end{center}
\end{footnotesize}
\end{table}

\revv{\subsubsection{Intermediate conclusions}
The Bouc-Wen system is a challenging case study since it inherently falls outside the PNLSS model class both for coupled as for decoupled models. The model structure allows only to approximate the absolute value nonlinearity by a finite number of terms. We have shown that the decoupled structure is a much more efficient parameterisation since it allows to increase the nonlinear degrees without having to suffer from a combinatorial growth of the number of parameters, which is the case for classical coupled polynomials. It was moreover demonstrated that the user can balance accuracy of the approximate model to complexity by ending the reduction when the performance falls below a given threshold.}
 
\subsection{An experimental battery model}
\label{ss:battery}

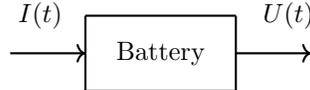
\begin{figure}[h]
\begin{center}
\begin{tikzpicture}
\draw [thick,->] (0.5,0) -- (1.5,0);
\draw [thick] (1.5,0) -- (1.5,-0.5);
\draw [thick] (1.5,0) -- (1.5,0.5);
\draw [thick] (1.5,0.5) -- (3.5,0.5);
\draw [thick] (1.5,-0.5) -- (3.5,-0.5);
\draw [thick] (3.5,-0.5) -- (3.5,0.5);
\draw [thick,->] (3.5,0) -- (4.5,0);
\node at (2.5,0) {Battery};
\node at(0.9,0.5) {$I(t)$};
\node at(4.2,0.5) {$U(t)$};
\end{tikzpicture}
\caption{\color{black} Single-input single-output Battery system. Current is considered the input and voltage is measured as the output.}
\label{f:battery_scheme}
\end{center}
\end{figure}

The use of batteries increased significantly over the last decade with the advent of electric cars and other means of electric transportation. To maximise the autonomy, the boundaries of the operating regimes have been stretched to low states-of-charge and potentially non-ideal temperature conditions. Under these circumstances batteries are known to respond nonlinearly. Typically a current to voltage relationship is sought. For such a type of system, building \revv{white-box} models is a difficult task. In \cite{relan2016} it was shown that \revv{black-box} PNLSS models are however capable of accurately capturing this relationship for a Li-Ion battery. \revv{The system is represented schematically in Fig.~\ref{f:battery_scheme}.}

\subsubsection{Data set}

As \textbf{training data}, a single realisation of an odd random-phase multisine is used. The band of excitation is between 1Hz and 5Hz, corresponding to the dynamic range of the battery. The excited spectrum has a resolution of $f_0=0.01$ Hz. A sample frequency of $f_s=50$ Hz was used. The data correspond to a state-of-charge (SoC) of 10\% and a temperature of $25^{\circ}$C.

\textbf{Validation} is carried out on an additional period of the training realisation.

\subsubsection{Results}

\begin{figure}
\begin{center}
\begin{subfigure}[b]{0.3\textwidth}
\begin{center}
\includegraphics[width=\textwidth]{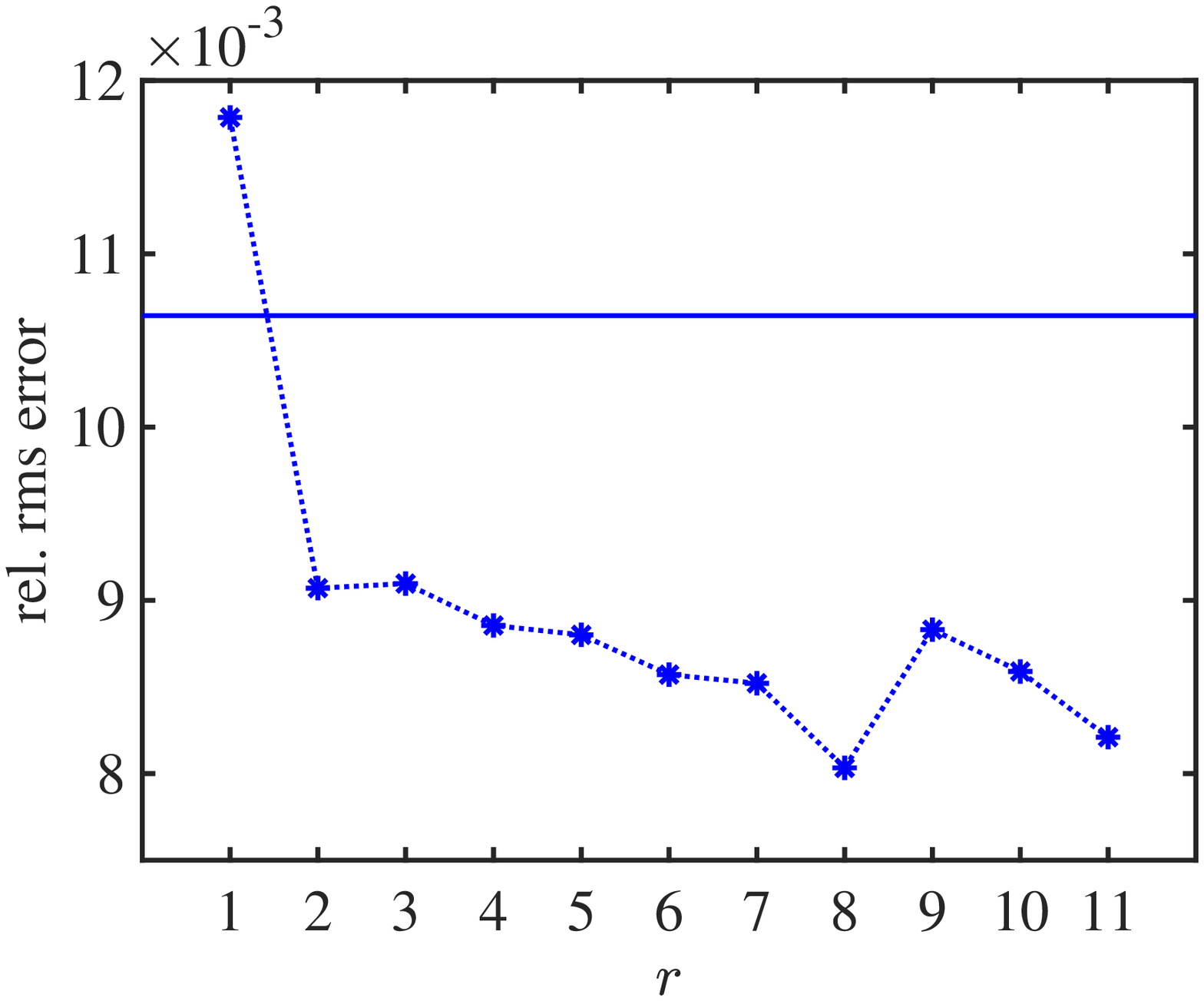}
\caption{}
\label{f:error_branches_battery}
\end{center}
\end{subfigure}
\begin{subfigure}[b]{0.3\textwidth}
\begin{center}
\includegraphics[width=\textwidth]{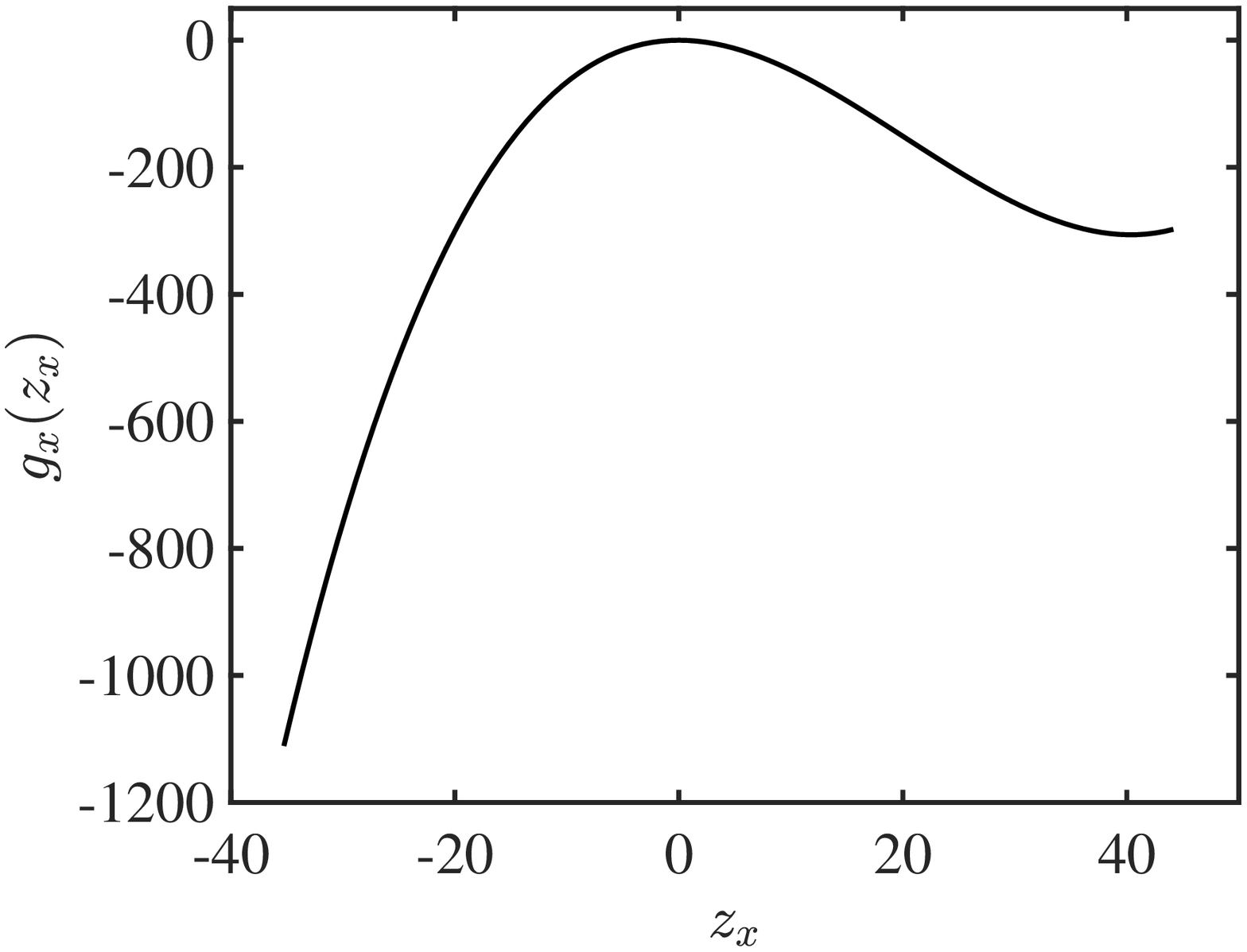}
\caption{}
\label{f:gx_battery}
\end{center}
\end{subfigure}
\begin{subfigure}[b]{0.3\textwidth}
\begin{center}
\includegraphics[width=\textwidth]{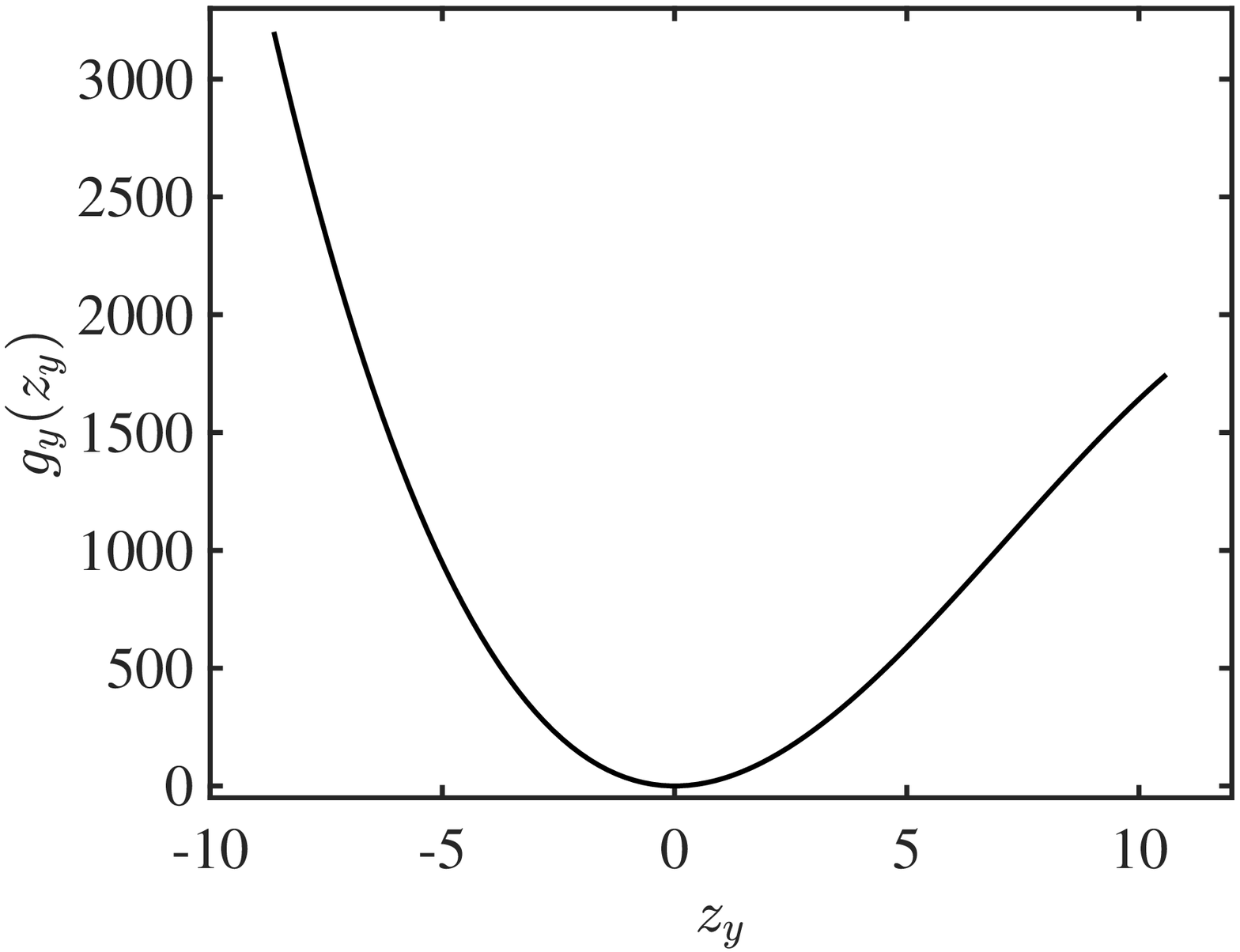}
\caption{}
\label{f:gy_battery}
\end{center}
\end{subfigure}
\caption{(a) Relative rms error on the validation data of the Li-Ion battery system as a function of the number of branches in the reduced model. Dots are the results of the reduced model while the solid line corresponds to the original PNLSS model. (b) Visualisation of the branch in the state equation for the $r=1$ model, evaluated on the training data. (c) Visualisation of the branch in the output equation for the $r=1$ model, evaluated on the training data.}
\label{f:val_battery}
\end{center}
\end{figure}

\begin{figure}
\begin{center}
\includegraphics[width=\textwidth]{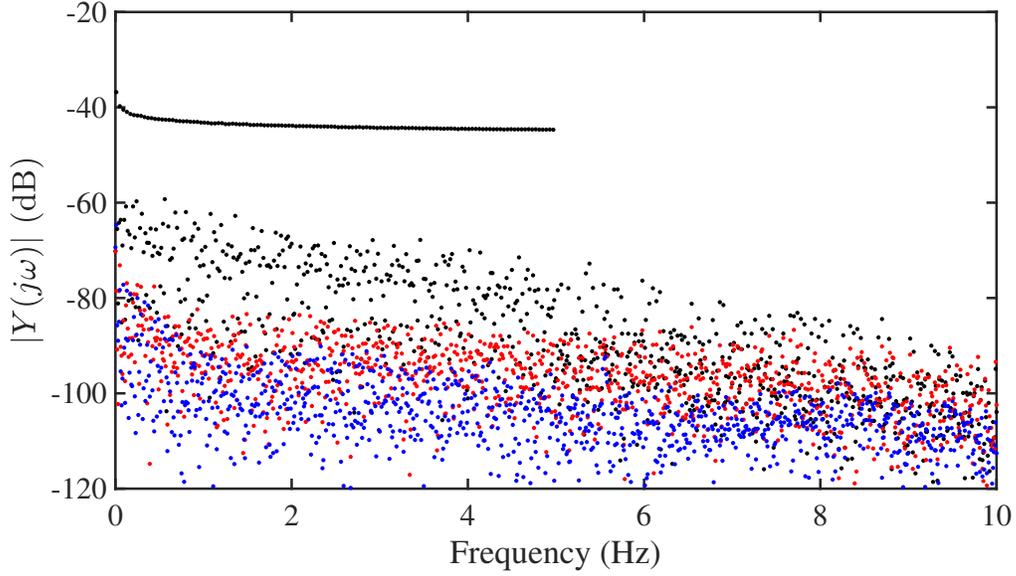}
\caption{Spectrum of the validation results of the 1-branch battery model.  Black shows the output data, blue is the error of the original coupled PNLSS model and red is the error of the 1-branch decoupled model.}
\label{f:val_R_battery}
\end{center}
\end{figure}

\rev{The PNLSS model, which is of fourth order \revv{($n=4$)}, contains both a state ($\textbf{f}_x$) and an output ($\textbf{f}_y$) nonlinearity. Following the rank estimation of the Jacobian tensors $\mathcal{J}_x$ and $\mathcal{J}_y$, \revv{resulting} independently from $\textbf{f}_x$ and $\textbf{f}_y$, the decoupling is computed for $r_x = 11$ and $r_y = 4$ branches. In a next step the decoupled functions are reduced. It was opted first to reduce the decoupled output nonlinearity to the point where $r_y=1$. Doing so the overall model contained less parameters to be tuned during the more crucial decoupling of the state nonlinearity.

The results of the reduction are summarised in Fig.~\ref{f:error_branches_battery} and additionally listed in Table~\ref{t:results_battery}. It can be concluded that the coupled PNLSS model can be outperformed by the decoupled formulations for all values of $r_x >1$. The performance of the $r_x=1$, $r_y=1$ model is slightly worse although still comparable to the PNLSS reference. The final model corresponds to a reduction of the used degrees of freedom from 259 to only 22. 
 


The remaining branches in the state and the output equation (visualised in Fig.~\ref{f:gx_battery} and Fig.~\ref{f:gy_battery}) appear to be dominantly \revv{even functions on the training domain. Even nonlinearities point to asymmetries in charging-discharging cycles.} The spectrum of the validation data is shown in Fig.~\ref{f:val_R_battery}.}

\begin{table}[h!]
\caption{Model reduction results of the Li-Ion battery system.}
\label{t:results_battery}
\begin{center}
\begin{tabular}{| c | c | c |}
\cline{2-3}
\multicolumn{1}{c|}{} & coupled PNLSS & $r=1$  \\
\hline
number of inputs to the nonlinearity & 5 & 5 \\
state nonlinearity $\textbf{f}_x$ &  degrees 2,3 & degrees 2, 3\\
output nonlinearity $\textbf{f}_y$ &  degrees 2,3 & degrees 2, 3\\
\# DOF & 259 & 22 \\
$e_{\text{rms}}$ validation realisation & 0.0106 & 0.0118\\
\hline
\end{tabular}
\end{center}
\end{table}

\revv{\subsubsection{Intermediate conclusions}
One of the challenges of black-box identification is that the complexity of the underlying system is generally unknown. This results in the pitfall of identifying complex models to inherently low-complex systems. The latter is observed for the battery case study. In this case the linear model was estimated to be of the fourth order. Since the number of terms in the classical coupled polynomial grows combinatorially with both the degree and the number of inputs, having four state variables results in bulky functions while the nonlinearity can in fact be grasped accurately by single branch functions.}

\subsection{Open problems: a model of unsteady fluid dynamics}
\label{ss:VIV}

In this section the limitations of the presented method are illustrated on the basis of an experimental case study.

In \cite{decuyper2018} the forces that arise on a submerged cylinder in a uniform flow were modelled. The fluctuating forces originate back from an unsteady wake, characterised by alternating vortices. The phenomenon is of particular interest in civil engineering given the frequent encounter of slender, cylindrical shapes in the built environment.

An accurate PNLSS model that relates the displacement of the cylinder to the resulting fluid forces was derived. Attempts to decouple this model have so far been unsuccessful. The specifics of the PNLSS model are listed in Table~\ref{t:VIV_results}.

\begin{table}[h!]
\begin{center}
\caption{Specifics of the PNLSS model of unsteady fluid dynamics}
\label{t:VIV_results}
\begin{tabular}{| c | c |}
\cline{2-2}
\multicolumn{1}{c|}{} & coupled PNLSS\\
\hline
state nonlinearity $\textbf{f}_x$ &  degrees 0-7\\
output nonlinearity $\textbf{f}_y$ &  degrees 0-7\\
\# DOF & 1396 \\
\hline
\end{tabular}
\end{center}
\end{table}

Two main challenges are encountered: 
\begin{itemize}
\item The dimensions of $\textbf{f}_x$ and $\textbf{f}_y$ are very large, resulting in a large number of branches. To ensure smoothness (see Section \ref{ss:avoiding_pr}) on all branches a dedicated regularisation penalty must be derived for the individual branches.
\item The PNLSS model is very sensitive to instability \rev{(in the sense that bounded input levels may result in unbounded output levels)}. Small decoupling errors may render the model unstable on the training data, making any further optimisation steps unfeasible.
\end{itemize}

This problem will be solved by implementing a more robust optimisation method that is able to deal with unstable systems.

\revv{\subsection{Interpreting the nonlinearity}
We have shown that models can be significantly reduced in size using the decoupling and reduction techniques. Apart from the benefits of having a smaller model, e.g.\ lower computation time, lower demand for storage space, one would also welcome an improved understanding of the system under test. Even after reduction, the model remains of the black-box type. Interpretation is therefore a hard problem. In some cases, however, the simplicity of the one-branch structure can be exploited. The model may then be examined with the aim of providing physical interpretation. Can we, for example, classify the type of nonlinear behaviour.

\subsubsection{Classifying the nonlinearity: a ``mechanical" interpretation}
\label{ss:class}
Insight is a subtle notion, especially for nonlinear systems where intuition is hard to develop. What can aid the understanding is the ability to relate the behaviour to something familiar. From a mechanical perspective one may want to classify a nonlinear system as being either a nonlinear spring, a nonlinear damper or both. Analogously, a nonlinear impedance and a nonlinear inductor can be considered. What discriminates both nonlinear components is the `input' on which the nonlinearity acts. Terms of the form of $f(y(t))$, with $f$ a nonlinear function and $y(t)$ the output of the model, can be perceived as springs since they store potential energy while terms of the form of $f(\dot{y}(t))$ are related to dampers. 

One-branch models contain a single univariate function as description of the nonlinearity. Examining their single input variable $z(t)$ (see e.g.\ Eq.~\eqref{e:SB2a}) to the nonlinear function allows to classify the behaviour as a damping or spring-like effect. Essential is a decomposition of $z(t)$ into $y(t)$ and $\dot{y}(t)$ components,
\begin{equation}
\label{e:z_recon}
\textbf{z} = [\textbf{y} \quad \dot{\textbf{y}}] \bm{\theta}_z,
\end{equation}
where $\textbf{z} \in \mathbb{R}^{L-1}$ is the vector of input samples to the nonlinear branch, $\textbf{y}  \in \mathbb{R}^{L-1}$ is the vector of output samples of the model and $\dot{\textbf{y}} \in \mathbb{R}^{L-1}$ is computed from \textbf{y} using a finite difference approximation. The data record length is denoted by $L$. Solving Eq.~\eqref{e:z_recon} for $\bm{\theta}_z \in \mathbb{R}^{2}$ decomposes the signal.\footnote{Technically, $\ddot{\textbf{y}}$ could be added to the regressor in Eq.~\eqref{e:z_recon}, although it is less common for mechanical systems to have nonlinear inertia.}
\begin{itemize}
\item If \textbf{z} is decomposable, i.e.\ it can be reconstructed from the proposed regressor matrix in Eq.~\eqref{e:z_recon}, the orientation of the $\bm{\theta}_z$ vector serves as an indicator for damping and/or spring-like behaviour.
\item If \textbf{z} cannot be decomposed into such a form the test is inconclusive and it eludes on the fact that the underlying system does not admit a one-branch model.
\end{itemize}
Notice that $z(t)$ is an informative variable since even-though state-space models are non-unique under a state transformation (see Eq.~\eqref{e:PNLSS_T} in Section \ref{ss:def_PNLSS}), such transformations leave $z(t)$ unaltered. Consider the one-branch single-input single-output model
 \begin{subequations} \label{e}
    \begin{empheq}[left={\empheqlbrace\,}]{align}
      & \textbf{x}(k+1)=\textbf{A}\textbf{x}(k)+\textbf{b}u(k)+ \textbf{w}g(z(k)) \color{black} \\  
      & y(k)=\textbf{c}^{\text{T}}\textbf{x}(k)+du(k),\\
      & z(k) = \textbf{v}^{\textbf{T}} \textbf{x}(k),
\end{empheq}
\end{subequations}
with matrices of the following dimensions, $\textbf{A} \in \mathbb{R}^{n \times n}$ and $\{\textbf{b},\textbf{w}, \textbf{c}, \textbf{v}\} \in \mathbb{R}^n$. Introducing an invertible state transformation matrix $\textbf{T} \in \mathbb{R}^{n \times n}$, such that $\textbf{x}(k)=\textbf{T}\textbf{x}_T(k)$ results in
\begin{subequations} \label{e}
    \begin{empheq}[left={\empheqlbrace\,}]{align}
      & \textbf{x}_T(k+1)=\textbf{T}^{-1}\textbf{A}\textbf{T}\textbf{x}_T(k)+\textbf{T}^{-1}\textbf{b}u(k)+ \textbf{T}^{-1} \textbf{w}g(z_T(k)) \color{black} \label{e:SB2a}\\  
      & y(k)=\textbf{c}^{\text{T}}\textbf{T}\textbf{x}_T(k)+du(k),\\
      & z_T(k) = \textbf{v}^{\textbf{T}}\textbf{T} \textbf{x}_T(k),
\end{empheq}
\end{subequations}
from which by construction $z(k) = z_T(k)$ for all \textbf{T}. Note that this also holds in the more general case when $z(k) = \textbf{v}^{\text{T}} [\textbf{x}(k)~\textbf{u}(k)]^{\text{T}}$.

\begin{itemize}
\item \textbf{The forced Duffing oscillator}

The \textbf{z} vector is obtained from simulating the multisine validation data (see Section \ref{ss:silverbox}). In this case \textbf{z} can be decomposed into components \textbf{y} and $\dot{\textbf{y}}$ up to a precision of 99.2\% (1-$e_{\text{rms}}$). The $\bm{\theta}_z$ vector is reported normalised
\begin{equation}
\frac{\bm{\theta}_z}{\norm{\bm{\theta}_z}} = [1.0000 \quad -0.0001]^{\text{T}}.
\end{equation}
It classifies the system as a (nearly) pure nonlinear spring. Judging from the shape of the nonlinearity (Fig.~\ref{f:r1_SB}) it may be called a hardening spring.
\item \textbf{The forced Van der Pol oscillator}

The \textbf{z} vector is obtained from simulating the multisine validation data (see Section \ref{ss:VdP}). In this case, the decomposition of \textbf{z} into components \textbf{y} and $\dot{\textbf{y}}$ is accurate up to a precision of 99.8\%. The normalised $\bm{\theta}_z$ vector yields
\begin{equation}
\frac{\bm{\theta}_z}{\norm{\bm{\theta}_z}} = [0.9999 \quad -0.0135]^{\text{T}}.
\end{equation}
It classifies the system as both a nonlinear spring and a nonlinear damper. This was anticipated given the $y^2(t)\dot{y}(t)$ nonlinear term in Eq.\eqref{e:diff_VdP}. Fig.~\ref{f:r1_VdP} suggests hardening effects. The ratio between spring and damping behaviour is subjected to the operating regime present throughout the data used in simulating \textbf{z}.

\item \textbf{The Bouc-Wen system}

The \textbf{z} vector is obtained from simulating the multisine validation data (see Section \ref{ss:Bouc_Wen}). In this case \textbf{z} cannot be decomposed accurately into components \textbf{y} and $\dot{\textbf{y}}$ yielding a precision of only 78.0\% (1-$e_{\text{rms}}$). The test is therefore inconclusive. Recall that a one-branch model was not able to accurately describe the Bouc-Wen system. No conclusions can be drawn from decomposing the $z$'s in case of multiple branches since stiffness and damping effects may be canceled out by a linear combination in \textbf{W}.
\item \textbf{The Battery system}

The Battery model contains a nonlinear function in both the state and the output equation. Also here the test is inconclusive since there is more than one input signal to the nonlinearity.
\end{itemize}
}

\section{Conclusions}
\label{s:conclusions}

This article provides a method to tackle the complexity of \revv{black-box} polynomial nonlinear state-space (PNLSS) models. It was found that a more efficient parameterisation of the nonlinear elements of PNLSS models can be found compared to the generic multivariate polynomials which are \revv{classically} used. The method relies on a tensor decomposition of the first order derivate information in order to decouple the multivariate polynomials into a set of univariate functions. Additional reduction is provided by imposing constraints on the obtained set of functions and/or reducing their number. The reduced models contain nonlinear elements which are much easier to interpret. \revv{In some cases insight into the system is obtained, classifying it as a nonlinear spring, a nonlinear damper, or both}. During the successive reduction steps the accuracy of the model is monitored. Hence, the user is able to balance model complexity to accuracy. \revv{Two additional benefits of the decoupled form were illustrated: (1) the decoupling step alters the optimisation landscape, this may potentially result in convergence to a more accurate local optimum when compared to the coupled counterpart, and (2) in the decoupled form the number of parameters grows linearly with the degree of the function while it grows combinatorially for coupled polynomials.} The method was illustrated on the following numerical and experimental case studies: the forced Duffing oscillator, the forced Van der Pol oscillator,  the Bouc-Wen hysteresis model, and a Li-Ion battery model.

\section{Acknowledgements}
The authors would like to thank Philippe Dreesen for the valuable contributions that were made.

This work was supported by the Fund for Scientific Research (FWO-Vlaanderen), the Swedish Research Council (VR) via the project NewLEADS -- New Directions in Learning Dynamical Systems (contract number: 621-2016-06079), and by the Swedish Foundation for Strategic Research (SSF) via the project ASSEMBLE (contract number: RIT15-0012).

\newpage
\section*{References}
\bibliography{entirebib}

\begin{thebibliography}{10}
\expandafter\ifx\csname url\endcsname\relax
  \def\url#1{\texttt{#1}}\fi
\expandafter\ifx\csname urlprefix\endcsname\relax\def\urlprefix{URL }\fi
\expandafter\ifx\csname href\endcsname\relax
  \def\href#1#2{#2} \def\path#1{#1}\fi

\bibitem{pintelon2001}
R.~Pintelon, J.~Schoukens, System Identification: A Frequency Domain Approach.,
  IEEE Press, 2001.

\bibitem{bai2014}
E.~W. Bai, K.~Li, W.~Zhao, On variable selection of a nonlinear non-parametric
  system with a limited data set: A stepwise algorithm, in: Preprints of the
  19th World Congress The International Federation of Automatic Control, 2014.

\bibitem{billings2013}
S.~Billings, Nonlinear System Identification: {NARMAX} Methods in the Time,
  Frequency and Spatio-Temporal Domains, Wiley, 2013.

\bibitem{usevich2014}
K.~Usevich, Decomposing multivariate polynomials with structured low-rank
  matrix completion, in: 21st International Symposium on Mathematical Theory of
  Networks and Systems ({MTNS}), 2014, pp. 1826--1833.

\bibitem{schoukensM2012}
M.~Schoukens, Y.~Rolain, Cross-term elimination in parallel {W}iener systems
  using a linear input transformation, {IEEE} Transactions on Instrumentation
  and Measurement 61~(3) (2012) 845--847.

\bibitem{dreesen2014}
P.~Dreesen, M.~Ishteva, J.~Schoukens, Decoupling multivariate polynomials using
  first-order information, {SIAM} Journal on Matrix Analysis and Applications
  36~(2) (2014) 864--879.

\bibitem{decuyper2019}
J.~Decuyper, P.~Dreesen, J.~Schoukens, M.~Runacres, K.~Tiels, Decoupling
  multivariate polynomials for nonlinear state-space models, {IEEE} {C}ontrol
  {S}ystems {L}etters 3~(3) (2019) 745--750.

\bibitem{paduart2010}
J.~Paduart, L.~Lauwers, J.~Swevers, K.~Smolders, J.~Schoukens, R.~Pintelon,
  Identification of nonlinear systems using polynomial nonlinear state space
  models, Automatica 46 (2010) 647--657.

\bibitem{noel2017}
J.-P. No{\"e}l, A.~Fakhrizadeh~Esfahani, G.~Kerschen, J.~Schoukens, A nonlinear
  state-space approach to hysteresis identification, Mechanical Systems and
  Signal Processing 84 (2017) 171--184.

\bibitem{relan2016}
R.~Relan, Y.~Firouz, J.-M. Timmermans, J.~Schoukens, Data driven nonlinear
  identification of {L}i-ion battery based on a frequency domain nonparametric
  analysis, {IEEE} Transactions on {C}ontrol {S}ystems {T}echnology 25~(5)
  (2017) 1825--1832.

\bibitem{decuyper2018}
J.~Decuyper, T.~De~Troyer, M.~Runacres, K.~Tiels, J.~Schoukens, Nonlinear
  state-space modelling of the kinematics of an oscillating circular cylinder
  in a fluid flow, Mechanical Systems and Signal Processing 98 (2018) 209--230.

\bibitem{young2018}
P.~Young, A.~Janot, Efficient parameterisation of nonlinear system models: a
  comment on {No{\"e}l} and {Schoukens} (2018), International Journal of
  Control DOI: 10.1080/00207179.2018.1521008.

\bibitem{fakhrizadeh2018}
A.~Fakhrizadeh~Esfahani, P.~Dreesen, J.~P. No{\"e}l, K.~Tiels, J.~Schoukens,
  Parameter reduction in nonlinear state-space identification of hysteresis,
  {Mechanical Systems And Signal Processing} 104 (2018) 884.

\bibitem{dreesen2016}
P.~Dreesen, A.~Fakhrizadeh~Esfahani, J.~Stoev, K.~Tiels, J.~Schoukens,
  Decoupling nonlinear state-space models: case studies, in: Proceedings of the
  International Conference on Noise and Vibration Engineering (ISMA), 2016, pp.
  2639--2646.

\bibitem{wigren2013}
T.~Wigren, J.~Schoukens, Three free data sets for development and benchmarking
  in nonlinear system identification, in: European Control Conference ({ECC}),
  Zurich, Switzerland, 2013, pp. 2933--2938.

\bibitem{fakhrizadeh2018Phd}
A.~Fakhrizadeh~Esfahani, Structure discrimination and identification of
  nonlinear systems, Ph.D. thesis, Vrije Universiteit Brussel (2018).

\bibitem{tensorlab}
N.~Vervliet, O.~Debals, L.~Sorber, M.~Van~Barel, L.~De~Lathouwer, Tensorlab
  3.0, {http://tensorlab.net/} (2016).

\bibitem{kruskal1977}
J.~B. Kruskal, Three-way arrays: rank and uniqueness of trilinear
  decompositions, with application to arithmetic complexity and satistics, Lin.
  Algebra Appl. 18 (1977) 95--138.

\bibitem{kruskal1989}
J.~B. Kruskal, Rank decomposition, and uniqueness for 3-way and {N}-way arrays,
  Elsevier Science Publishers B.V., 1989.

\bibitem{hollander2018}
G.~Hollander, Multivariate polynomial decoupling in nonlinear system
  identification, Ph.D. thesis, Vrije Universiteit Brussel (2017).

\bibitem{dreesen2018}
P.~Dreesen, J.~De~Geeter, M.~Ishteva, Decoupling multivariate functions using
  second-order information and tensors, in: Proc. 14th International Conference
  on Latent Variable Analysis and Signal Separation {LVA/ICA} 2018, Vol. 10891,
  Springer, 2018, pp. 79--88.

\bibitem{sorber2015}
L.~Sorber, M.~Van~Barel, L.~De~Lathauwer, Structured data fusion, {IEEE}
  {J}ournal of {S}elected {T}opics in {S}ignal {P}rocessing 9~(4) (2015)
  586--600.

\bibitem{noel2018}
J.-P. No{\"e}l, J.~Schoukens, Grey-box state-space identification of nonlinear
  mechanical vibrations, International Journal of Control 91~(5) (2018)
  1118--1139.

\bibitem{ljung2004}
L.~Ljung, Q.~Zhang, P.~Lindskog, A.~Juditski, Modeling a non- linear electric
  circuit with black box and grey box models, in: Proc. IFAC Symposium on
  Nonlinear Control Systems {(NOLCOS2004)}, 2004, pp. 543--548.

\bibitem{sragner2004}
L.~Sragner, J.~Schoukens, G.~Horvath, Modeling of slightly nonlinear systems: a
  neural network approach, in: Proc. IFAC Symposium on Nonlinear Control
  Systems {(NOLCOS2004)}, 2004, pp. 531--536.

\bibitem{vanderpol1926}
B.~Van~der Pol, Relaxatie-trillingen, Tijdschrift van het Nederlandsch
  radiogenootschap 3 (1926) 25--40.

\bibitem{dowell1981}
E.~H. Dowell, Non-linear oscillator models in bluff body aero-elasticity,
  Journal of Sound and Vibration 75~(2) (1981) 251--264.

\bibitem{hartlen1970}
R.~T. Hartlen, I.~G. Currie, Lift-oscillator model of vortex-induced vibration,
  Journal of Engineering Mechanics Division 96~(5) (1970) 577--591.

\bibitem{parkinson1974}
G.~V. Parkinson, Mathematical models of fluid-induced vibrations of bluff
  bodies, Flow-induced structural vibrations., Springer, 1974.

\bibitem{morrison2001}
D.~Morrison, Y.~Jia, J.~Moosbrugger, Cyclic plasticity of nickel at low plastic
  strain amplitude: hysteresis loop shape analysis, Materials {S}cience and
  {E}ngineering A314 (2001) 24--30.

\bibitem{bertotti1998}
G.~Bertotti, Hysteresis in Magnetism, Academic Press, San Diego, 1998.

\bibitem{mueller1985}
T.~Mueller, The influence of laminar separation and transition on low
  {Reynolds} number airfoil hysteresis, {AIAA} Journal of Aircraft 22~(9)
  (1985) 763--770.

\bibitem{schoukens2017}
M.~Schoukens, J.-P. No{\"e}l, Three benchmarks addressing open challenges in
  nonlinear system identification, in: 20th World Congress of the International
  Federation of Automatic Control, Toulouse, France, 2017, pp. 448--453.

\end{thebibliography}

\end{document}